\newif\ifconfver
\confvertrue        

\newif\ifplainver  
\plainvertrue

\ifplainver
    \confverfalse   
\fi

\ifconfver
     \documentclass[10pt,twocolumn,twoside]{IEEEtran}
\else
    \ifplainver
        \documentclass[11pt]{article}
        \usepackage{fullpage}
    \else
        \documentclass[12pt,draftcls,onecolumn]{IEEEtran}
    \fi
\fi

\usepackage{calc,amsfonts,amssymb,amsmath,bm,url,color,theorem,graphicx,cite}
\usepackage{psfrag,subfigure,float}
\usepackage{algorithm}
\usepackage{algorithmic}
\usepackage{soul}
\usepackage{blkarray}

\definecolor{orange}{RGB}{255,107,0}

\newtheorem{Fact}{Fact}

\newtheorem{Prop}{Proposition}
\newtheorem{Theorem}{Theorem}

\newtheorem{Corollary}{Corollary}

\theorembodyfont{\rmfamily}

\newcommand\bw{\ensuremath{{\bm w}}}
\newcommand\bW{\ensuremath{{\bm W}}}
\newcommand\cW{\ensuremath{\bm{\mathcal{W}}}}

\newcommand\bx{\ensuremath{{\bm x}}}
\newcommand\by{\ensuremath{{\bm y}}}
\newcommand\bh{\ensuremath{{\bm h}}}
\newcommand\bH{\ensuremath{{\bm H}}}
\newcommand\cH{\ensuremath{\bm{\mathcal{H}}}}
\newcommand\bbh{\ensuremath{\bar{\bm h}}}

\newcommand\be{\ensuremath{{\bm e}}}
\newcommand\bX{\ensuremath{{\bm X}}}
\newcommand\bZ{\ensuremath{{\bm Z}}}
\newcommand\cZ{\ensuremath{\bm{\mathcal{Z}}}}
\newcommand\bC{\ensuremath{{\bm C}}}
\newcommand\bA{\ensuremath{{\bm A}}}
\newcommand\bb{\ensuremath{{\bm b}}}
\newcommand\blam{\ensuremath{{\bm \lambda}}}
\newcommand\bmu{\ensuremath{{\bm \mu}}}

\newcommand\bXi{\ensuremath{{\bm \Xi}}}
\newcommand\bPi{\ensuremath{{\bm \Pi}}}
\newcommand\bbPi{\ensuremath{\bar{\bm \Pi}}}
\newcommand\bF{\ensuremath{{\bm F}}}
\newcommand\bbF{\ensuremath{\bar{\bm F}}}
\newcommand\bv{\ensuremath{{\bm v}}}
\newcommand\bG{\ensuremath{{\bm G}}}

\newcommand{\Rbb}{\mathbb{R}}
\newcommand{\Cbb}{\mathbb{C}}
\newcommand{\Hbb}{\mathbb{H}}

\newcommand{\setU}{\mathcal{U}}
\newcommand{\setV}{\mathcal{V}}
\newcommand{\setS}{\mathcal{S}}
\newcommand{\setC}{\mathcal{C}}
\newcommand{\setE}{\mathcal{E}}

\newcommand{\setI}{\mathcal{I}}
\newcommand{\setX}{\mathcal{X}}
\newcommand{\setY}{\mathcal{Y}}

\newcommand\bQ{\ensuremath{{\bm Q}}}
\newcommand\br{\ensuremath{{\bm r}}}

\newcommand{\eps}{\varepsilon}
\newcommand{\bzero}{{\bm 0}}
\newcommand{\bone}{{\bm 1}}
\newcommand{\bI}{{\bm I}}
\newcommand*{\LargerCdot}{\raisebox{-0.25ex}{\scalebox{1.2}{$\cdot$}}}

\begin{document}

\bibliographystyle{IEEEtran}

\newcommand{\papertitle}{
Unraveling the Rank-One Solution Mystery of Robust MISO Downlink Transmit Optimization: A Verifiable Sufficient Condition via a New Duality Result}

\newcommand{\paperabstract}{
This paper concentrates on a robust transmit optimization problem for the multiuser multi-input single-output (MISO) downlink scenario and under inaccurate channel state information (CSI).
This robust problem deals with a general-rank transmit covariance design and follows a safe rate-constrained formulation under spherically bounded CSI uncertainties.
Curiously,  simulation results in previous works suggested that the robust problem admits rank-one optimal transmit covariances in most cases.
Such a numerical finding is appealing
because transmission with rank-one covariances can be easily realized by single-stream transmit beamforming.
This gives rise to a fundamentally important question,
namely,
whether we can theoretically identify conditions under which the robust problem admits a rank-one solution.
In this paper, we 
identify
one such condition.
Simply speaking, 
we show that
the robust problem is guaranteed to admit a rank-one solution
if the CSI uncertainties are not too large and the multiuser channel is not too poorly conditioned.
To establish the aforementioned condition,
we develop a novel duality framework, through which an intimate relationship between the robust problem and a related maximin problem is revealed.
Our condition involves only a simple expression with respect to the multiuser channel and other system parameters.
In particular, unlike other sufficient rank-one conditions that have appeared in the literature, ours is verifiable.
The application of our analysis framework to several other CSI uncertainty models is also discussed.
}


\ifplainver


    \title{\papertitle}

    \author{
    Wing-Kin Ma$^\dag$, Jiaxian Pan$^\dag$, Anthony Man-Cho So$^\ddag$, and Tsung-Hui Chang$^\ast$ \\ ~ \\
    $^\dag$Department of Electronic Engineering, \\
    The Chinese University of Hong Kong, Shatin, N.T., Hong Kong S.A.R. of China \\
    Email: \texttt{\{wkma,jxpan\}@ee.cuhk.edu.hk}
    \\ ~ \\
    $^\ddag$Department of Systems Engineering and Engineering Management,  \\
    The Chinese University of Hong Kong, Shatin, N.T.,  Hong Kong S.A.R. of China \\
    Email: \texttt{manchoso@se.cuhk.edu.hk}
    \\ ~ \\
    $^\ast$School of Science and Engineering, \\
    The Chinese University of Hong Kong, Shenzhen,  China \\
    Email: \texttt{tsunghui.chang@ieee.org}
    }

    \maketitle

    \begin{abstract}
    \paperabstract
    \end{abstract}

\else
    \title{\papertitle}

    \ifconfver \else {\linespread{1.1} \rm \fi

    \author{Wing-Kin Ma, Jiaxian Pan, Anthony Man-Cho So, and Tsung-Hui Chang
    \thanks{Wing-Kin Ma is with the Department of Electronic Engineering, The Chinese University of Hong Kong (CUHK), Hong Kong; email: \texttt{wkma@ee.cuhk.edu.hk.}
    Jiaxian Pan is with MediaTek Inc., Hsinchu, Taiwan; email: \texttt{jxpan@ee.cuhk.edu.hk.}
    Anthony Man-Cho So is with the Department of Systems Engineering and Engineering Management, CUHK, Hong Kong; email: \texttt{manchoso@se.cuhk.edu.hk.}
    Tsung-Hui Chang is with the School of Science and Engineering, CUHK, Shenzhen, P.R. China; email: \texttt{tsunghui.chang@ieee.org.}
    This work was supported in part by a Direct Grant of CUHK, Project ID 4055009.
    }
    
    \ifconfver
     \else
    \ifplainver
     \else
        \thanks{Copyright (c) 2016 IEEE. Personal use of this material is permitted. However, permission to use this material for any other purposes must be obtained from the IEEE by sending a request to \texttt{pubs-permissions@ieee.org}.}
    \fi
    \fi
    
    }

    \maketitle

    \ifconfver \else
        \begin{center} \vspace*{-2\baselineskip}
        \end{center}
    \fi

    \begin{abstract}
    \paperabstract
    \\\\
    \end{abstract}

    
    \ifconfver
    \else
    \vspace{2em}
     \noindent \small {\bfseries EDICS}: 
     OPT-CVXR (Convex optimization and relaxation for SP),
     SPC-INTF (Interference management techniques)
     \fi

    \ifconfver \else \IEEEpeerreviewmaketitle} \fi

 \fi

\ifconfver \else
    \ifplainver \else
        \newpage
\fi \fi

\section{Introduction}

In the multiuser multi-input multi-output (MIMO) downlink scenario,
linear precoding has played a key role in greatly enhancing system throughput and efficiency~\cite{BK:Bengtsson01,spencer2004zero,peel2005vector,gesbert2007shifting,gershman2010cvx_bf,bjornson2014lecturenote,baligh2014cross}.
In simple terms, the idea is to 
share the channel among multiple users simultaneously by leveraging on the MIMO degrees of freedom.
Linear precoding achieves this by
transmitting linearly superimposed multiuser signals
whose mutual interference, or multiuser interference, at the user side has been pre-managed by the base station.
However,
linear precoding also requires the base station to have access to the channel state information (CSI) of the users, most preferably perfect, 
for otherwise it would be difficult to perform precise interference control.
While 
it has been demonstrated in the literature (e.g., the above referenced articles)
that 
linear precoding can boost the per-user and overall achievable rates drastically in the perfect CSI regime, 
it is also well known that existing systems often do {\em not} acquire CSI perfectly
owing to a variety of practical and operational reasons~\cite{love2008overview}.
This practical constraint has stimulated a branch of research
that aims
to establish linear precoder design, or transmit optimization, frameworks that are robust against CSI uncertainties~\cite{Shenouda2007,ZhengWongNg_2008,Zheng_etal2009,Boche2009}.

It would not be easy to have a complete overview on the present developments of robust multiuser MIMO transmit optimization.
The reason is that existing works may use different system settings, and their technical developments usually have much dependence on the latter.
Particularly, a study can differ in terms of 
i) the chosen quality-of-service (QoS) measure for the users,
e.g., achievable rate, signal-to-interference-and-noise ratio, or symbol mean squared error;
ii) structural assumptions on the linear precoder,
e.g., 
general linear precoding with arbitrary-rank transmit covariances,
or
transmit beamforming with a fixed number of data streams;
iii) the design criterion,
e.g., the QoS-constrained formulation, or the sum-rate maximizing formulation.
Despite such
diversity, 
we can classify existing works into three types according to the robust performance metric.
The first is the worst-case approach, wherein the CSI uncertainties are seen as bounded deterministic unknowns (e.g., within a sphere), and the robust performance metric is the worst-case QoS with respect to (w.r.t.) the CSI uncertainties.
This leads to a transmit solution that is ``safe'' in the worst-case sense.
In this context the design criteria 
usually give rise to robust optimization problems with semi-infinite constraints.
The current state of the art focuses mainly on the conic optimization framework,
where techniques originated from robust optimization are applied to convert those semi-infinite constraints into (convex) linear matrix inequalities.
Note that
 the conversion is sometimes equivalent~\cite{Boche2009}, 
 and sometimes approximate as a restriction~\cite{Shenouda2007,tajer2011robust,huang2013lorentz} or a relaxation~\cite{ZhengWongNg_2008,Zheng_etal2009};
that generally depends on the chosen QoS.

The second type of robust techniques is the average approach, wherein the CSI uncertainties are modeled as random variables, 
and the average QoS w.r.t. the CSI uncertainties is used as the robust performance metric.
This average approach may lead to higher throughput than the worst-case approach in an average sense,
although one should also note that average robust performance measures may not be as desirable for real-time or delay-sensitive traffic.
The corresponding design criteria 
lead to stochastic optimization problems, which present a different challenge;
some recent results can be found in 
\cite{RazaviyaynSSUM13,Yang2016parallel}.
The third type is the outage-based approach,
wherein the performance metric is 
a QoS level under which the actual QoS will be satisfied with high probability
(again assuming that the  CSI uncertainties are random).
It is a safe approach, but in a probabilistic sense and with the level of pessimism being tunable.
Outage-based designs deal with chance-constrained optimization problems that are intrinsically very hard to solve.
Recent developments tackle this issue through efficiently computable approximations~\cite{Shenouda2008,Shenouda13,sohrabi2016coordinate,wang2014outage,qli2014chance,he2015tight}.
It is interesting to note that the worst-case approach, upon appropriate modifications, can also be used to handle the outage-based designs; see, e.g., \cite{wang2014outage}.

\subsection{Focus of this Paper and Contribution}

In this paper we focus on a particular robust transmit optimization problem in the multiuser multi-input single-output (MISO) downlink scenario.
Specifically, the performance metric is the worst-case achievable rate under a spherically bounded CSI uncertainty model, and under general linear precoding.
The design criterion is that of minimizing the total transmission power, subject to the constraint that the worst-case achievable rate of each user is better than or equal to a pre-specified rate value.
The same problem was first studied in \cite{ZhengWongNg_2008} 
in the context of semidefinite relaxation (SDR)-based transmit beamforming and 
extended later to other scenarios such as cognitive radio networks~\cite{Zheng_etal2009}, distributed multicell coordination~\cite{shen2012distributed}, and outage-based robust designs \cite{wang2014outage}.
In particular, the problem can be converted to a semidefinite program (SDP), which can be efficiently solved by conic optimization algorithms.
Empirically, it has been observed that this robust problem exhibits a very desirable behavior, namely, 
the optimal transmit covariances of the multiuser signals were found to be of rank one in most of the instances~\cite{ZhengWongNg_2008,Zheng_etal2009,wang2014outage}.
We should stress that the design criterion does not impose any rank constraints on the transmit covariances,
and numerical results 
suggested 
that rank-one optimal transmit covariances is generally the case.
This phenomenon is practically important,
since in rank-one instances 
the physical-layer transmit strategy reduces to (per-user) single-stream transmit beamforming, which is simple to deploy in practice.

The contribution of this paper is fundamental.
We 
set out to prove when the robust transmit optimization problem described above admits a rank-one solution.
From a communication theory viewpoint, our motivation is similar to that in some classical MIMO study, such as the single-user average robust study in \cite{visotsky2001space}, where the objective is to understand when the simple single-stream transmit beamforming scheme is the optimal physical-layer transmit scheme.
Moreover, from a mathematical optimization perspective,
this rank-one solution analysis problem is closely related to the study of 
rank reduction theory in SDP,
which is important as evidenced in 
recent works \cite{AsurveyoftheSlemma,SYZ08,SY10,Sagnol11,nagy2014forbidden}.

Before we describe our approach, let us mention some related work.
Under the same system setting, rank-one solution analyses in the perfect CSI case have been considered, and in fact solved, in \cite{BK:Bengtsson01,Huang10TSP}.
The provable rank-one results therein are strong and requires little assumption.
However, the main tools used in the perfect CSI case, such as the SDP rank reduction technique~\cite{Huang10TSP}, turn out to be not too useful in the 
inaccurate CSI case;
this will be discussed in the next section.
In that regard, we are faced with a new analysis challenge.
Recently,
several independent studies have attempted to solve the robust rank-one solution analysis problem~\cite{Song11,Chang2011,wang2013tightness}
by identifying various sufficient conditions for the robust problem to admit a rank-one solution.
Unfortunately, these sufficient conditions are not verifiable in the sense that they either depend on some quantity that 
cannot be determined in closed form
or require certain assumptions whose satisfiability has not been further proven.
Thus,  
it is not easy to extract physical meanings from those results.  
As an additional minor note,
the robust problem is shown to have strong rank-one solution guarantees in certain restrictive cases, namely, when there is only one user~\cite{shen2012distributed,zheng2009robust}, or when there are at most two transmit antennas~\cite{Song11}.

In this work, we prove a verifiable sufficient condition for the robust rank-one solution analysis problem.  
Roughly speaking, we show that if the magnitudes of the CSI uncertainties are small compared to those of the corresponding channels,
and if the channels of different users are not too similar in terms of direction,
then the robust problem will admit a rank-one solution.
The aforementioned condition sounds practically reasonable, 
since large CSI uncertainties and similar channel directions tend to result in infeasibility of the robust problem or a poor solution in terms of power efficiency.
Our analysis is based on a 
novel duality result,
proven herein specifically for the robust problem.
This duality result allows us to tackle the robust rank-one solution analysis question by studying the rank-one solution conditions of the same problem under perfect CSI, which is an arguably easier task.
The duality result also provides fundamentally new insights into the robust problem, as we will explain in the paper.

\subsection{Organization and Notations}

In Section~\ref{sec:bg} we give the background of the robust transmit optimization problem of interest.
This will also include simulation results and a concise review of some known rank-one solution results.
In Section~\ref{sec:main} we describe our main rank-one result.
This is followed by Section~\ref{sec:proof}, which provides the proof of the main result.
Section~\ref{sec:ext} discusses how the main result can be applied to several other CSI uncertainty models.
Section~\ref{sec:con} concludes the paper.

The notations used in this paper are mostly standard, and some specific notations are defined as follows:
$\Hbb^n$ is the set of all complex-valued Hermitian $n \times n$ matrices;
$\bX \succeq \bzero$ and $\bX \succ \bzero$ mean that $\bX$ is positive semidefinite and positive definite, respectively;
$\bx \geq \bzero$ means that $\bx$ is elementwise nonnegative;
$\| \bx \|_2 = \sqrt{ \bx^H \bx }$ and $\| \bx \|_\infty  = \max_{i=1,\ldots,n} | x_i |$ are the $2$-norm and infinity-norm, respectively;
${\mathbb{E}}[ \cdot ] $ denotes expectation;
the superscript $\dag$ means the pseudo-inverse.


\section{Background Review}
\label{sec:bg}


\subsection{System Model}

Consider a unicast multiuser MISO downlink scenario, in which the base station transmits information signals, one for each user, simultaneously.
The signal transmitted by the base station is given by $\bx(t)= \sum_{i=1}^K \bx_i(t)$, where $\bx_i(t) \in \Cbb^N$ is the vector information signal for user $i$, $K$ is the number of users, 
and $N$ is the number of 
antennas
at the base station.
Also, each $\bx_i(t)$ is generated independently from one another.
Every channel from the base station to a user is assumed to be frequency-flat and static within the transmission time block. 
Correspondingly, the received signal of user $i$, $i = 1,\ldots,K$, is $y_i(t)= \bh_i^H \bx(t) + v_i(t)$,
where $\bh_i \in \Cbb^N$ is the channel from the base station to user $i$,
and $v_i(t)$ is complex circular Gaussian noise with mean zero and variance $\sigma_i^2$.
By denoting the transmit covariance of $\bx_i(t)$ as $\bW_i = {\mathbb{E}}[ \bx_i(t) \bx_i^H(t)] \in \Hbb^N$ and assuming vector-Gaussian signaling for every $\bx_i(t)$,
the achievable rate of user $i$ can be modeled as  
\[
{\sf R}_i( \cW,  \bh_i) = \log_2 \left( 1 + \frac{ \bh_i^H \bW_i \bh_i }{ \sum_{j \neq i}  \bh_i^H \bW_j \bh_i + \sigma_i^2 } \right),
\]
where, for 
conciseness, we denote $\cW = (\bW_1,\ldots, \bW_K)$.

We deal with transmit optimization, or the design of $\cW$ for enhancing system performance.
In this context,
a crucial assumption is that the base station has acquired the CSI $( \bh_i )_{i=1}^K$.
How the CSI is acquired is system-dependent.  For instance, in frequency-division duplex (FDD) systems we use quantized channel feedback, while in time-division duplex (TDD) systems we use uplink channel estimation.
In general, we can write
\[
\bh_i = \bbh_i + \be_i, \quad i=1,\ldots,K,
\]
where $\bbh_i$ is the presumed or estimated channel of user $i$ at the base station, and $\be_i$ represents the corresponding channel error.
The channel errors capture uncertainties caused by a combination of several effects, 
such as time variations of the channels before and after channel acquisition,
channel quantization errors in FDD, 
and channel estimation errors (which depend on a number of factors in the underlying physical-layer structures;
e.g., in LTE, those factors include the OFDM resource block structures, the corresponding pilot assignment scheme, and the channel estimation algorithm employed).
An accurate model for such a channel error process can be complicated and system-dependent.  For this reason, simple but effective models are usually preferred.
One such model is the spherically bounded model,
where $\be_i$'s are treated as deterministic unknowns with
\[ \| \be_i \|_2   \leq \eps_i, \quad i=1,\ldots,K, \]
where $\eps_i > 0$ represents a known worst-case error magnitude bound.

\subsection{The Robust Rate-Constrained Problem}
\label{sec:RRC}

Under the above system setup, the transmit optimization problem of interest is to minimize the total transmission power
and make sure 
every user will achieve a rate that is no less than a pre-specified value under any spherically bounded channel uncertainties.
Mathematically, this is formulated as
\begin{equation} \label{eq:main_rate}
\begin{aligned}
\min_{ \bW_1, \ldots, \bW_K \in \Hbb^N} & ~ \textstyle \sum_{i=1}^K {\rm Tr}(\bW_i)   \\
{\rm s.t.} \qquad & ~  {\sf R}_i(\cW, \bh_i) \geq r_i ~ \text{for all $\bh_i \in \setU_i$},  
~ i=1,\ldots,K, \\
& ~ \bW_1, \ldots, \bW_K \succeq \bzero,
\end{aligned}
\end{equation}
where 
$r_i > 0$ is the pre-specified rate value of user $i$,
and 
\[ \setU_i = \{ \bh_i \in \Cbb^N ~|~ \| \bh_i - \bbh_i \|_2 \leq \eps_i \} \]
denotes the admissible channel set for user $i$.
Problem~\eqref{eq:main_rate} is a robust transmit covariance design following the worst-case approach.
For convenience, Problem~\eqref{eq:main_rate} will be called the {\em robust rate-constrained problem} in the sequel.

The robust rate-constrained problem is a ``good'' transmit optimization problem in the sense that it can be solved using conic optimization machinery.
To see this, let
\begin{align*}
\varphi_i(\cW,\bh_i) & = \sigma_i^2 + \bh_i^H \left(  \sum_{j \neq i} \bW_j  - \frac{1}{\gamma_i} \bW_i \right) \bh_i, \\
\gamma_i & = 2^{r_i} - 1 > 0
\end{align*}
for $i=1,\ldots,K$, and observe that the rate constraint ${\sf R}_i(\cW, \bh_i) \geq r_i$ can be rewritten as $\varphi_i(\cW,\bh_i) \leq 0$.
Thus, Problem~\eqref{eq:main_rate} can be expressed as
\begin{subequations} \label{eq:main}
\begin{align}
\min_{ \cW} & ~ \textstyle \sum_{i=1}^K {\rm Tr}(\bW_i)   \\
{\rm s.t.} & ~  \max_{\bh_i \in \setU_i} \varphi_i(\cW,\bh_i) \leq 0,  ~ i=1,\ldots,K,  \label{eq:main_b} \\
& ~ \bW_1, \ldots, \bW_K \succeq \bzero. \label{eq:main_c}
\end{align}
\end{subequations}
Note that
we leave the assumption $\bW_i \in \Hbb^N$, $i=1,\ldots,K$, implicit for notational conciseness, and
the same convention will be applied hereafter.
Since $\varphi_i$ is affine in $\cW$, Problem \eqref{eq:main} is convex. 
However, the constraints in \eqref{eq:main_b} are semi-infinite.
Fortunately,
in this particular case 
such constraints can be easily tackled.
The idea is to apply the $\mathcal{S}$-lemma 
\cite{AsurveyoftheSlemma}:
Since $\setU_i$ and $\varphi_i(\cW, \bh_i)$ are quadratically dependent on $\bh_i$,
the $\mathcal{S}$-lemma implies that 
the constraints in  \eqref{eq:main_b} can be equivalently transformed into the
linear matrix inequalities (LMIs)
\begin{equation*}
\exists t_i \geq 0 \text{~such that~}
\begin{bmatrix}
\bQ_i + t_i \bI & \br_i \\
\br_i^H & \ s_i - t_i \eps_i^2 
\end{bmatrix} \succeq \bzero,
~i=1,\ldots,K,
\end{equation*}
where $\bQ_i = \textstyle \frac{1}{\gamma_i} \bW_i - \sum_{j \neq i} \bW_j$,
$\br_i = \bQ_i \bbh_i$,  $s_i = \bbh_i^H  \bQ_i \bbh_i - \sigma_i^2$;
see \cite{ZhengWongNg_2008}.
Plugging the above LMIs into Problem~\eqref{eq:main}, we 
can reformulate Problem~\eqref{eq:main} as
\begin{subequations}  \label{eq:main_sdp}
\begin{align}
\min_{ \cW, \cZ, {\bm t} } & ~ \textstyle \sum_{i=1}^K {\rm Tr}(\bW_i)   \\
{\rm s.t.} \,\,\,& ~  
\bZ_i= \begin{bmatrix}
\bQ_i + t_i \bI & \br_i \\
\br_i^H & \ s_i - t_i \eps_i^2 
\end{bmatrix}, 
 ~i=1,\ldots,K,
 \label{eq:main_sdp_lmi}
 \\
& 
~ \bW_i \succeq \bzero, \bZ_i \succeq \bzero, t_i \geq 0, ~i=1,\ldots,K.
\end{align}
\end{subequations}
The above problem is an SDP and can be efficiently solved by available conic optimization software~\cite{sedumi,cvx}.
Note that the aforementioned SDP formulation 
was first proposed in \cite{ZhengWongNg_2008}.

\subsection{The Rank-One Solution Mystery}

There is an interesting phenomenon, and also a mystery, concerning the robust rate-constrained problem.
It has been observed by numerical means that the optimal solution $\cW^\star$ to Problem~\eqref{eq:main} 
is almost always
of rank one
(i.e., ${\rm rank}(\bW_i^\star)= 1$ for all $i$),
and this was consistently reported in several independent studies such as \cite{ZhengWongNg_2008,Song11,Chang2011,wang2014outage}.
Such a result is very meaningful,
since in the rank-one case we can easily realize the achievable rates in physical layer via single-stream beamforming; specifically,
$\bx_i(t) = \bw_i s_i(t)$, 
where $\bw_i \in \Cbb^N$ is a beamforming vector and $s_i(t)$ is a zero-mean unit-power data stream for user $i$ (note also the equivalence  $\bW_i = \bw_i \bw_i^H \Longleftrightarrow {\rm rank}(\bW_i) \leq 1, \bW_i \succeq \bzero$).
As an additional remark, one can also find physical-layer transceiver schemes that are designed to handle higher-rank $\bW_i^\star$ (see, e.g., 
\cite{wu2012rank,wen2012rank,wu2013SBF,law2013general,wustochastic2}), 
but 
single-stream beamforming is simpler to implement than those schemes.
Readers are referred to \cite{ZhengWongNg_2008,wang2014outage} for further descriptions concerning the SDR interpretation of Problem~\eqref{eq:main} for single-stream beamforming design.

To give readers some insight,
we produce a set of test results in Table~\ref{tab:rank1}.
We see that except for some very occasional instances,
Problem~\eqref{eq:main} has a rank-one solution whenever it is feasible.
We should note that higher-rank instances were also spotted in the ellipsoidal channel error model; see the examples shown in \cite{Zheng_etal2009,Song11}.
Nevertheless, those instances are arguably rare.

\ifconfver
	\begin{table*}[thb]
\else
	\begin{table}[hbt]
\fi
\caption{Occurrence of rank-one solution to the robust rate-constrained problem in \eqref{eq:main}.
The test was conducted on $2,000$ randomly generated channel instances,
and we
 set $r_1 = \cdots = r_K = r$ and $\sigma_i^2= 0.1$.
} 
\vspace{1em}
\centering
\ifconfver
\else
  \tiny
  \setlength{\tabcolsep}{4pt}
\fi
    \begin{tabular}{r|c|c|c|c|c|c|c|c|c|c}
          & \multicolumn{10}{c}{number of rank-1 instances / number of feasible instances} \\
\cline{2-11}    \multicolumn{1}{c|}{$r$} & \multicolumn{2}{c|}{$(N,K)=(4,3)$} & \multicolumn{2}{c|}{$(N,K)=(8,3)$} & \multicolumn{2}{c|}{$(N,K)=(8,7)$} & \multicolumn{2}{c|}{$(N,K)=(12,7)$} & \multicolumn{2}{c}{$(N,K)=(12,11)$} \\
\cline{2-11}    \multicolumn{1}{c|}{(bits/s/Hz)} & $\eps_i^2=0.1$ & $\eps_i^2=0.05$ & $\eps_i^2=0.1$ & $\eps_i^2=0.05$ & $\eps_i^2=0.1$ & $\eps_i^2=0.05$ & $\eps_i^2=0.1$ & $\eps_i^2=0.05$ & $\eps_i^2=0.1$ & $\eps_i^2=0.05$ \\
    \hline
    \hline
    \multicolumn{1}{c|}{0.1375} & 2000/2000 & 2000/2000 & 2000/2000 & 2000/2000 & 2000/2000 & 2000/2000 & 2000/2000 & 2000/2000 & 2000/2000 & 2000/2000 \\
    \multicolumn{1}{c|}{0.2122} & 2000/2000 & 2000/2000 & 2000/2000 & 2000/2000 & 2000/2000 & 2000/2000 & 2000/2000 & 2000/2000 & 2000/2000 & 2000/2000 \\
    \multicolumn{1}{c|}{0.3233} & 2000/2000 & 2000/2000 & 2000/2000 & 2000/2000 & 2000/2000 & 2000/2000 & 2000/2000 & 2000/2000 & 2000/2000 & 2000/2000 \\
    \multicolumn{1}{c|}{0.4835} & \textbf{1999/2000} & 2000/2000 & 2000/2000 & 2000/2000 & 2000/2000 & 2000/2000 & 2000/2000 & 2000/2000 & 2000/2000 & 2000/2000 \\
    \multicolumn{1}{c|}{0.7057} & \textbf{1999/2000} & 2000/2000 & 2000/2000 & 2000/2000 & 2000/2000 & 2000/2000 & 2000/2000 & 2000/2000 & 2000/2000 & 2000/2000 \\
    \multicolumn{1}{c|}{1.0000} & 1973/1973 & 1995/1995 & 2000/2000 & 2000/2000 & 2000/2000 & 2000/2000 & 2000/2000 & 2000/2000 & 2000/2000 & 2000/2000 \\
    \multicolumn{1}{c|}{1.3701} & 1933/1933 & 1993/1993 & 2000/2000 & 2000/2000 & 2000/2000 & 2000/2000 & 2000/2000 & 2000/2000 & 2000/2000 & 2000/2000 \\
    \multicolumn{1}{c|}{1.8122} & 1688/1688 & 1889/1889 & 2000/2000 & 2000/2000 & \textbf{1950/1952} & 1997/1997 & 2000/2000 & 2000/2000 & 2000/2000 & 2000/2000 \\
    \multicolumn{1}{c|}{2.3165} & 1535/1535 & 1833/1833 & 2000/2000 & 2000/2000 & 1084/1084 & 1814/1814 & 1999/1999 & 2000/2000 & \textbf{1483/1485} & 1976/1976 \\
    \multicolumn{1}{c|}{2.8698} & 1258/1258 & 1743/1743 & 2000/2000 & 2000/2000 &  271/ 271 &  995/ 995 & 1964/1964 & 1998/1998 &  109/ 109 & 1068/1068 \\
    \multicolumn{1}{c|}{3.4594} &  839/ 839 & 1539/1539 & 1994/1994 & 2000/2000 &   51/  51 &  549/ 549 & 1795/1795 & 1993/1993 &    6/   6 &  160/ 160 \\
    \multicolumn{1}{c|}{4.0746} &  365/ 365 & 1187/1187 & 1961/1961 & 2000/2000 &    4/   4 &  181/ 181 & 1262/1262 & 1936/1936 &    0/   0 &   28/  28 \\
    \multicolumn{1}{c|}{4.7070} &   68/  68 &  688/ 688 & 1753/1753 & 1987/1987 &    0/   0 &   19/  19 &  354/ 354 & 1659/1659 &    0/   0 &    2/   2 \\
    \multicolumn{1}{c|}{5.3509} &    1/   1 &  211/ 211 &  955/ 955 & 1920/1920 &    0/   0 &    0/   0 &   12/  12 &  885/ 885 &    0/   0 &    0/   0 \\
    \multicolumn{1}{c|}{6.0022} &    0/   0 &   21/  21 &  106/ 106 & 1485/1485 &    0/   0 &    0/   0 &    0/   0 &  122/ 122 &    0/   0 &    0/   0 \\
    \multicolumn{1}{c|}{6.6582} &    0/   0 &    0/   0 &    1/   1 &  469/ 469 &    0/   0 &    0/   0 &    0/   0 &    0/   0 &    0/   0 &    0/   0 \\
    \end{tabular}%

\vspace{1em}
\label{tab:rank1}
\ifconfver
	\end{table*}
\else
	\end{table}
\fi

It is also interesting to benchmark some existing state-of-the-art methods, particularly those that 
consider (per-user) single-stream beamforming under the same formulation as Problem~\eqref{eq:main}.
We use the max-min-fair (MMF) rate
\ifconfver
    \begin{align*}
    r^\star_{\sf MMF} = \max\{ r ~|~ & \textstyle \min_{\bh_i \in \setU_i} {\sf R}_i(\cW, \bh_i) \geq r  ~ \forall i, ~
    \bW_i \succeq \bzero ~ \forall i,  ~ \\
    &  \textstyle \sum_{i=1}^K {\rm Tr}(\bW_i) \leq 
    P_{\sf tot} 
    \}
    \end{align*}
\else
    \[
    r^\star_{\sf MMF} = \max\{ r ~|~ \textstyle \min_{\bh_i \in \setU_i} {\sf R}_i(\cW, \bh_i) \geq r  ~ \forall i, ~
    \bW_i \succeq \bzero ~ \forall i,  ~ \sum_{i=1}^K {\rm Tr}(\bW_i) \leq 
    P_{\sf tot} 
    \}
    \]
\fi
as the performance metric for comparison, 
where $P_{\sf tot}$
is given and describes the total transmission power limit.
The MMF rate can be computed 
by using Problem~\eqref{eq:main}; see the bisection search in \cite{WieselTSP06} for details.
Also,
if the MMF rate with single-stream beamforming is desired, one can replace Problem~\eqref{eq:main} with its single-stream beamforming 
version; i.e,
\begin{equation} \label{eq:main_r1}
\begin{aligned}
\min_{ \bw_1,\ldots, \bw_K \in \Cbb^N } 
& ~ \textstyle \sum_{i=1}^K \| \bw_i \|_2^2   \\
{\rm s.t.} \quad\,\,\,\,\, & ~  \max_{\bh_i \in \setU_i} \varphi_i( (\bw_i \bw_i^H )_{i=1}^K,\bh_i) \leq 0,  ~ i=1,\ldots,K,
\end{aligned}
\end{equation}
which is obtained by substituting $\bW_i = \bw_i \bw_i^H$ into Problem~\eqref{eq:main}.
We illustrate in Fig.~\ref{fig:mmf}  the MMF rates of various methods
and in Table~\ref{tab:runtime} the running times.
In particular,
``RSDP'' refers to the application of Problem~\eqref{eq:main},
``RMMSE'' the robust minimum-mean-square-error method in \cite{Boche2009},
`RSOCP1''  the robust second-order cone programming (SOCP) method in \cite[Algorithm~3]{Shenouda2007},
``RSOCP2'' the robust SOCP method in \cite[Theorem~4]{tajer2011robust},
and ``RSOCP3'' the robust SOCP method in \cite{huang2013lorentz}.
Note that other than ``RSDP'', all the benchmarked methods are 
convex restrictive approximations of
the single-stream beamforming 
problem in \eqref{eq:main_r1}.
Moreover, in this numerical experiment, we found that ``RSDP'' gives rank-one solutions all the time.
From Fig.~\ref{fig:mmf} and Table~\ref{tab:runtime}, we observe that ``RSDP'', or Problem~\eqref{eq:main}, is most competitive in both MMF rate and runtime performance.

\ifconfver
	\begin{figure}[hbt]
\else
	\begin{figure}[htb]
\fi	
\begin{center}
\ifconfver
	{\resizebox{.5\textwidth}{!}{\includegraphics{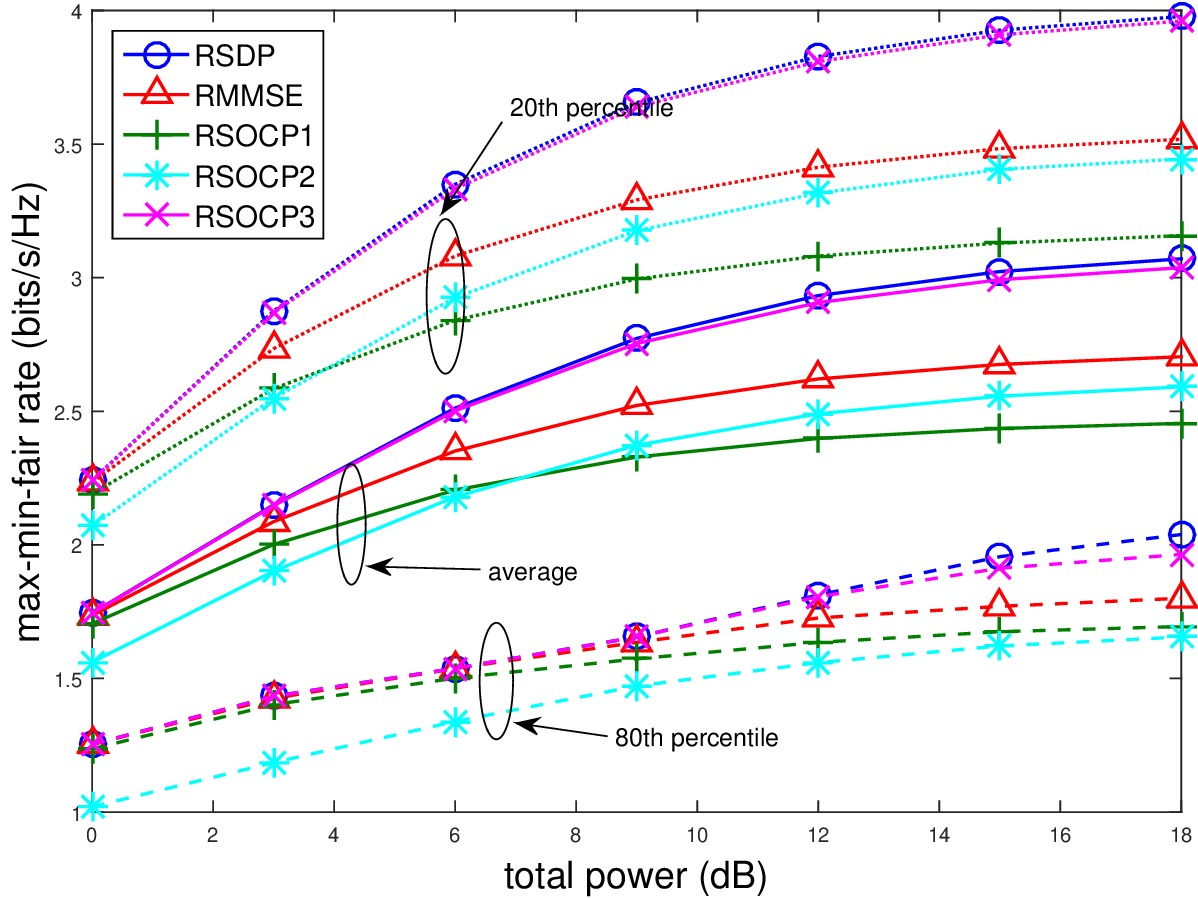}}}
\else
	{\resizebox{.75\textwidth}{!}{\includegraphics{fig1.eps}}}
\fi
\end{center}
\caption{Rate performance of various robust transmit optimization methods. $N=4$, $K= 3$, $\sigma_i^2= 0.1$, $\eps_i^2= 0.1$,
$2,000$ simulation trials.
The presumed channels are randomly generated at each trial, following a complex circular Gaussian distribution with mean $\bzero$ and covariance $\bI$.}
\label{fig:mmf}
\ifconfver
	\end{figure}
\else
	\end{figure}
\fi


\begin{table}[hbt]
\caption{Average runtime performance of the various methods for handling Problem~\eqref{eq:main} or the single-stream beamforming version of \eqref{eq:main}.
$N= 4$, $K= 3$, $\sigma_i^2= 0.1$, $\eps_i^2= 0.1$, $r_1= \cdots = r_K = 1.8122$, $2,000$ simulation trials.
\texttt{CVX}~\cite{cvx} was used to implement the various methods,
and the experiment was run on a PC with a CPU speed of $3.40$GHz.
}
\vspace{1em}
\centering
\begin{tabular}{c||c|c|c|c|c}
                       & RSDP & RMMSE & RSOCP1 & RSOCP2 & RSOCP3 \\ \hline
 time (in sec.) & 0.3577 & 0.3695   & 3.2419    & 0.5670     & 6.9603
\end{tabular}
\label{tab:runtime}
\vspace{1em}
\end{table}

\subsection{Problem Statement and Known Results}

The aforementioned numerical finding is very interesting and has motivated the following question:

\medskip
\noindent
{\bf Question:} \
{ \em
Under what conditions on the problem instance $\{ \bbh_i, \sigma_i^2, \eps_i \}_{i=1}^K$ will the robust rate-constrained problem \eqref{eq:main} admit a rank-one solution?
In addition, what are the 
subsequent
implications from a practical viewpoint;
e.g.,
requirements on the presumed channels 
$\bbh_i$'s,
limits on the uncertainty bounds $\eps_i$'s,
etc.?
}
\medskip

To better understand this analysis challenge, we start by reviewing the perfect CSI case.
From Problem \eqref{eq:main},
the rate-constrained problem under perfect CSI (or $\eps_i= 0$ for all $i$) is
\begin{equation} \label{eq:main_pcsi}
\begin{aligned}
\min_{ \cW} & ~ \textstyle \sum_{i=1}^K {\rm Tr}(\bW_i)   \\
{\rm s.t.} 
& ~  \textstyle {\rm Tr}( \bH_i ( \frac{1}{\gamma_i} \bW_i - \sum_{j \neq i} \bW_j ) ) \geq \sigma_i^2,  ~ i=1,\ldots,K,   \\
& ~ \bW_1, \ldots, \bW_K \succeq \bzero,
\end{aligned}
\end{equation}
where $\bH_i = \bh_i \bh_i^H$.
The above problem is a complex-valued separable SDP with $K$ separable decision variables and $K$ linear constraints.  For such a problem, it is well known that a rank-one solution exists;
e.g.,
by the Bengtsson-Ottersten uplink-downlink duality result (the first reported rank-one result in unicast transmit optimization) \cite{BK:Bengtsson01}, or by the SDP rank reduction technique \cite{Pataki98,Huang10TSP}.
Let us briefly review the latter approach by recalling a popularly used 
result:

\begin{Fact}[SDP Rank Reduction \cite{Huang10TSP}] \label{fac:rank_reduce}
Consider the complex-valued separable SDP
\begin{subequations} \label{eq:sdp_gen}
\begin{align}
\min_{ \bX_1, \ldots, \bX_k \in \Hbb^n}  & ~ \textstyle  \sum_{i=1}^k {\rm Tr}(\bC_i \bX_i)  \\
{\rm s.t.} \quad\,\,\,\,\,& ~  \textstyle   \sum_{l=1}^k {\rm Tr}(\bA_{i,l} \bX_l) \trianglerighteq_i b_i, ~i=1,\ldots,m,
\label{eq:sdp_gen_con} \\
& ~ \bX_1,\ldots,\bX_k \succeq \bzero,
\end{align}
\end{subequations}
where $\bA_{i,l}, \bC_i \in \Hbb^n$, $b_i \in \Rbb$ for all $i,l$,
and the notation $\trianglerighteq_i$ can be either `$\geq$' or `$=$' for each $i$.
Suppose that 
Problem~\eqref{eq:sdp_gen} has an optimal solution.\footnote{As a subtle point, in the literature SDP rank reduction results are usually proved under the assumption that Problem~\eqref{eq:sdp_gen} and its dual both have optimal solutions and attain zero duality gap. This assumption may be relaxed to that of only requiring Problem~\eqref{eq:sdp_gen} to have an optimal solution, through a variation of the existing 
proof;
see Theorem 5.4 and Corollary 5.5 in \cite{lemon2016low}.
}
Then, there exists a solution $(\bX_1^\star,\ldots,\bX_k^\star)$ to Problem~\eqref{eq:sdp_gen} such that
$$
\sum_{i=1}^k {\rm rank}(\bX_i^\star)^2\leq m. 
$$
In particular, if $\bX_i^\star \neq \bzero$ for all $i$ and $m \leq k+2$, then every $\bX_i^\star$ has ${\rm rank}(\bX_i^\star)=1$.
\end{Fact}
Armed with Fact~\ref{fac:rank_reduce}, one can easily conclude that Problem~\eqref{eq:main_pcsi} has a rank-one solution.

Thus,
one would be tempted to see whether SDP rank reduction can also be applied to 
the robust rate-constrained problem.
Unfortunately, this approach appears to have fundamental limitations.
In Appendix~\ref{apx:sdp_rank_reduce},
we prove that a direct application of Fact~\ref{fac:rank_reduce} leads to the following result:
If Problem~\eqref{eq:main_sdp} has an optimal solution,
then
there exists an optimal solution $(\bW_i^\star, \bZ_i^\star, t_i^\star)_{i=1}^K$ to Problem~\eqref{eq:main_sdp} such that
\begin{equation} 
\label{eq:rank_bnd_robust2}
\sum_{i=1}^K {\rm rank}(\bW_i^\star)^2 \leq K(N^2+2N) -  \sum_{i=1}^K {\rm rank}(\bZ_i^\star)^2.
\end{equation}
Furthermore, every $\bZ_i^\star$ must satisfy ${\rm rank}(\bZ_i^\star) \leq N$.
Let us assume for the sake of argument that ${\rm rank}(\bZ_i^\star) = N$ for all $i$, which is the best case one can hope for.
The above bound then becomes
\[ \sum_{i=1}^K {\rm rank}(\bW_i^\star)^2 \leq 2NK, \]
which is still too loose to provide a meaningful result for ${\rm rank}(\bW_i^\star)= 1$ for all $i$. 

Song {\em et al.} \cite{Song11} have recently proven 
some rank-one results for the robust rate-constrained problem.
Rather than using SDP rank reduction, they studied the Karush-Kuhn-Tucker (KKT) conditions of the SDP formulation \eqref{eq:main_sdp}.
In particular, they proved the following result:
Suppose that Problem~\eqref{eq:main_sdp} is feasible, and let $v^\star$ denote the optimal objective value of Problem~\eqref{eq:main_sdp}.
If 
\begin{equation} \label{eq:song}
\eps_i^2 < \frac{\gamma_i \sigma_i^2}{v^\star},  \quad i=1,\ldots,K,
\end{equation}
then any solution $\cW^\star$ to Problem~\eqref{eq:main_sdp} must have ${\rm rank}(\bW_i^\star)=1$ for all $i$.
Physically, this implies that the robust rate-constrained problem should have a rank-one solution for sufficiently small uncertainty bounds $\eps_i$'s.
While insightful, the above result has a fundamental drawback---the optimal value $v^\star$ also depends on the problem instance $\{ \bbh_i, \sigma_i^2, \eps_i \}_{i=1}^K$.
It is not clear how $v^\star$ would scale with these parameters.

\section{Main Result}
\label{sec:main}

In this section we present our main result.
Let $\bbF= [~ \bbh_1,\ldots, \bbh_K ~]$ be the presumed multiuser channel matrix,
$\bbF_{-i}$ be a submatrix of $\bbF$ obtained by removing the $i$th column of $\bbF$,
and $\bar{\bPi}_i = \bI - \bbF_{-i} (\bbF^H_{-i} \bbF_{-i})^\dag \bbF^H_{-i}$ be the orthogonal complement projector of $\bbF_{-i}$.
The following theorem summarizes the result.

\begin{Theorem} \label{thm:main}
Suppose that Problem~\eqref{eq:main} has an optimal solution,
and that $\sigma_i^2 > 0$ for all $i$.
If
\begin{equation} \label{eq:main_cond}
\frac{ \| \bbPi_k \bbh_k \|_2^2 }{ \eps_k^2 } > 1+K + \left( K - \frac{1}{K} \right) \gamma_k,  \quad k=1,\ldots,K
\end{equation} 
holds,
then the optimal solution $\cW^\star$ to Problem~\eqref{eq:main} must have 
${\rm rank}(\bW_i^\star) =1$ for all $i$.
\end{Theorem}

Note that Theorem~\ref{thm:main} is a sufficient condition, and 
as discussed previously empirical experience indicates a much better situation.
A numerical result is shown in Fig.~\ref{fig:thm1} to give more insights.
There, we randomly generated many instances of $\bbh_i$'s and evaluated the empirical satisfaction probability of \eqref{eq:main_cond}.
As seen, there is generally a gap between the satisfaction probability of \eqref{eq:main_cond} and the probability that the robust problem is feasible.
That said, for $(N,K)= (12,3)$, the gap is almost zero when the rate requirement is below $3$~bits/s/Hz.
As a reference, Fig.~\ref{fig:thm1} also shows the satisfaction probability of the sufficient rank-one condition \eqref{eq:song} by Song {\em et al.}~\cite{Song11}.

\ifconfver
	\begin{figure}[hbt]
\else
	\begin{figure}[htb]
\fi	
\begin{center}
\ifconfver
	{\resizebox{.45\textwidth}{!}{\includegraphics{fig2_new.eps}}}
\else
	{\resizebox{.75\textwidth}{!}{\includegraphics{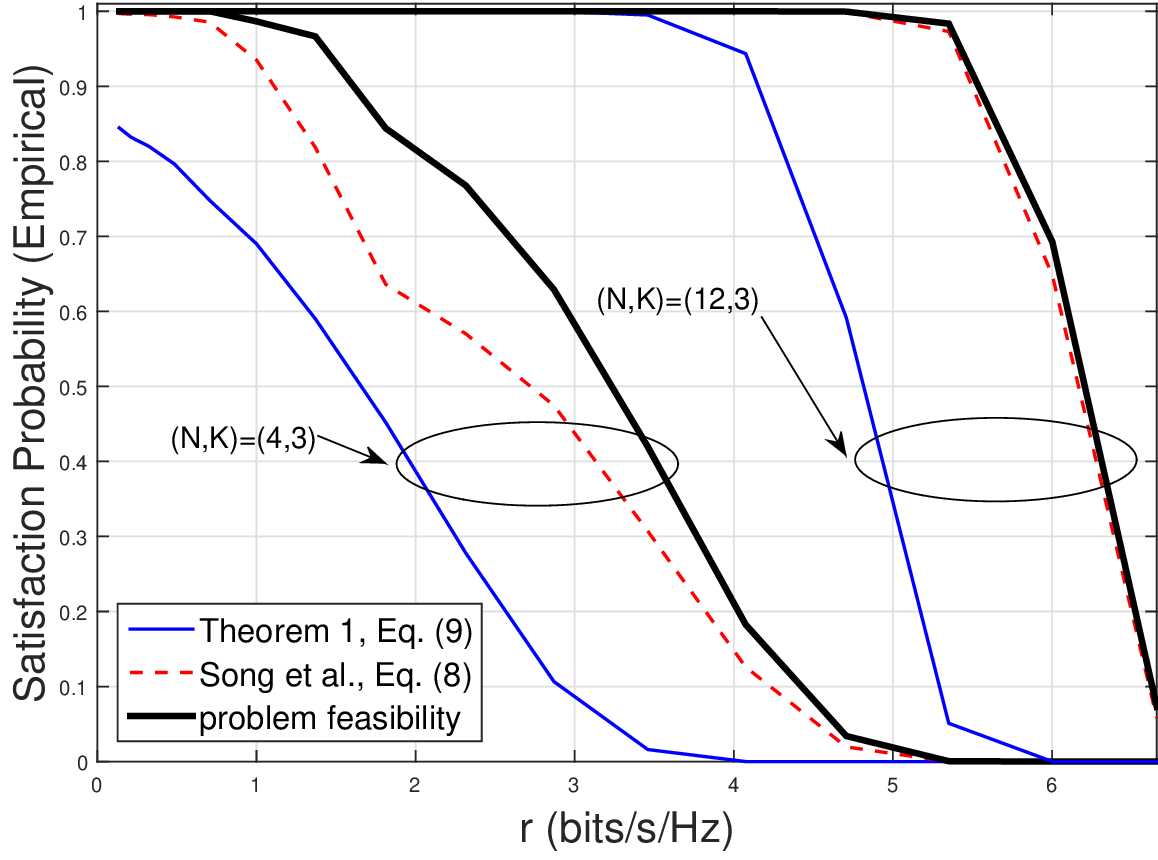}}}
\fi
\end{center}
\caption{The satisfaction probability of the sufficient rank-one condition in Theorem~\ref{thm:main}. 
$\sigma_i^2= 0.1$, $\eps_i^2= 0.1$,
$r_1 = \cdots =r_K = r$,
$2,000$ simulation trials.
The presumed channels are randomly generated at each trial, following a complex circular Gaussian distribution with mean $\bzero$ and covariance $\bI$.
}
\label{fig:thm1}
\ifconfver
	\end{figure}
\else
	\end{figure}
\fi

The proof of Theorem~\ref{thm:main} will be the focus in the next section.
Here, we are interested 
in extracting physical meanings from Theorem~\ref{thm:main}.
It is easy to verify that for $N \geq K$, we have
\begin{equation} \label{eq:main_imp0}
\| \bbPi_k \bbh_k \|_2 \geq \| \bbh_k \|_2 \cdot \sigma_{\rm min}(\hat{\bF}),
\end{equation}
where $$\hat{\bF}= [~ \bbh_1/\| \bbh_1 \|_2, \ldots, \bbh_K /\| \bbh_K \|_2 ~]$$ is 
the presumed multiuser channel direction matrix, 
and $\sigma_{\rm min}(\cdot)$ denotes the smallest singular value of its argument;
see Appendix~\ref{apx:eq:main_imp0} for the proof of \eqref{eq:main_imp0}.
We see that, as a direct corollary of Theorem~\ref{thm:main}, if $N \geq K$ and
\begin{equation} \label{eq:main_imp}
 \frac{\| \bbh_k \|_2^2}{\eps_k^2} \sigma_{\rm min}(\hat{\bF})^2 > 1+K + \left( K - \frac{1}{K} \right) \gamma_k,  \quad k=1,\ldots,K
\end{equation}
holds, then the optimal solution to Problem~\eqref{eq:main} must be of rank one.
The inequality~\eqref{eq:main_imp} has several implications.
First, fixing $K$ and $\gamma_k$'s,
the rank-one condition can be guaranteed if the channel-to-uncertainty ratios $\| \bbh_k \|_2^2 / \eps_k^2$ are sufficiently large and 
the presumed multiuser channel direction matrix
$\hat{\bF}$ is sufficiently well-conditioned.
Second, the rank-one condition becomes harder to satisfy if the number of users $K$ is larger and/or if the rate requirements $r_i$'s are higher (recall $\gamma_i = 2^{r_i}-1$).
Third, the rank-one condition does not depend on the noise powers $\sigma_i^2$'s.

We also have the following result:
\begin{Prop} \label{prop:prob}
Suppose that $\bbh_1,\ldots,\bbh_K$ are independent circularly symmetric complex Gaussian random vectors, where $\bbh_i$ has mean $\bzero$ and covariance $\rho_i \bI$ for some $\rho_i > 0$.
Let
\[
{\sf CUR}_i = \frac{ \mathbb{E}[ \| \bbh_i \|_2^2 ] }{ \eps^2_i } = \frac{ \rho_i N }{ \eps_i^2 }
\]
be the ratio of the $i$th presumed channel's average magnitude to the worst-case channel magnitude, or the channel-to-uncertainty ratio (CUR) for short.
Furthermore, let
\[
\eta_i = \frac{N}{N-K+1} \left[ 1+K + \left( K - \frac{1}{K} \right) \gamma_i \right],
\quad i=1,\ldots,K.
\]
If
${\sf CUR}_i > \eta_i$ for $i=1,\ldots,K$,
then the sufficient rank-one condition \eqref{eq:main_cond} in Theorem~\ref{thm:main} holds with probability at least
\[
1 - \sum_{k=1}^K \left( \frac{ \eta_k e }{ {\sf CUR}_k } \right)^{N-K+1},
\]
where $e\approx2.71828$ is the Euler number.
\end{Prop}
We relegate the proof of Proposition~\ref{prop:prob} to Appendix~\ref{app:prop:prob}.
Proposition~\ref{prop:prob} not only shows that the probability of admitting a rank-one robust solution is guaranteed to increase if the CURs increase but also indicates that such probability increases rapidly to $1$ as we make $N-K$ larger. In fact, the latter is in good agreement with the numerical result in Fig.~\ref{fig:thm1}.

\medskip
\noindent
{\bf Remark 1:} \
The condition in \eqref{eq:main_cond} can be improved in some cases.
In the proof of Theorem~\ref{thm:main}, we also show that if $ \| \bbPi_k \bbh_k \|_2^2 / \eps_k^2 \geq (K+1)^2$ for some $k$, the corresponding bound in  \eqref{eq:main_cond} may be replaced by 
\begin{equation} \label{eq:main_cond_var1}
\frac{ \| \bbPi_k \bbh_k \|_2^2 }{ \eps_k^2 } > ( 1 + \sqrt{ (K-1) \gamma_k } )^2.
\end{equation}


\section{Proof of Theorem~\ref{thm:main}}
\label{sec:proof}

The proof of Theorem~\ref{thm:main} contains two parts.
In the first part, we develop a duality-based analysis framework for the robust rate-constrained problem~\eqref{eq:main}.
Our analysis framework is quite different from those of Song {\em et al.} \cite{Song11}
and our previous preliminary work~\cite{Chang2011}, 
in the sense that  we do not rely on the $\mathcal{S}$-lemma-based SDP formulation \eqref{eq:main_sdp}.
The duality result also provides some interesting implications, which we will discuss.
In the second part, we prove a specific rank-one solution condition for the non-robust rate-constrained problem,
which, when applied to our duality framework, leads to the robust rank-one solution condition in Theorem~\ref{thm:main}.


\subsection{Preliminaries}

Let us describe the two basic ingredients in our proof.
The first is about properties of the fixed-channel, or perfect-CSI, rate-constrained problem. 
From \eqref{eq:main_pcsi}, consider
\begin{equation} \label{eq:main_pcsi2}
\begin{aligned}
\min_{ \cW} & ~ \textstyle \sum_{i=1}^K {\rm Tr}(\bW_i)   \\
{\rm s.t.} 
& ~  \textstyle {\rm Tr}( \bH_i ( \frac{1}{\gamma_i} \bW_i - \sum_{j \neq i} \bW_j ) ) \geq \sigma_i^2,  ~ i=1,\ldots,K,   \\
& ~ \bW_1, \ldots, \bW_K \succeq \bzero,
\end{aligned}
\end{equation}
where 
every $\bH_i$ is a general Hermitian matrix,
as opposed to $\bH_i = \bh_i \bh_i^H$ in the previous study.
The dual of Problem~\eqref{eq:main_pcsi2} is
\begin{equation} \label{eq:main_pcsi2_dual}
\begin{aligned}
\max_{ \bmu} & ~ \textstyle \sum_{i=1}^K \sigma_i^2 \mu_i   \\
{\rm s.t.} 
& ~  \textstyle \bI + \sum_{j \neq i} \mu_j \bH_j - \frac{\mu_i}{\gamma_i} \bH_i \succeq \bzero,  ~ i=1,\ldots,K,   \\
& ~ \mu_1, \ldots, \mu_K \geq 0,
\end{aligned}
\end{equation}
where $\mu_1, \ldots, \mu_K \in \Rbb$ are the dual variables.
For convex problems such as the above, strong duality and existence of optimal solutions should not be an issue in practice,
though their validity requires a proof.
A result arising from one such proof is as follows:

\begin{Fact} \label{fac:sd_pcsi}
Consider 
Problem~\eqref{eq:main_pcsi2} with
$\sigma_i^2 > 0$ for all $i$.
Suppose that Problem~\eqref{eq:main_pcsi2} is feasible.
Then, Problem~\eqref{eq:main_pcsi2} is also strictly feasible.
Moreover, its dual problem~\eqref{eq:main_pcsi2_dual} is strictly feasible 
regardless of the feasibility of Problem~\eqref{eq:main_pcsi2}.
Consequently, by the strong duality theorem (e.g., \cite{BK:Ben-Tal01}), 
Problems~\eqref{eq:main_pcsi2}--\eqref{eq:main_pcsi2_dual} both have optimal solutions, and they attain 
zero duality gap.
\end{Fact}

Fact~\ref{fac:sd_pcsi} can be easily proven.\footnote{
Concisely, if $\cW'$ is feasible to \eqref{eq:main_pcsi2},
then $\bW''_i = \alpha \bW_i' + \bI$, $i=1,\ldots,K$, is strictly feasible for some sufficiently large $\alpha > 1$.
Moreover, we can always find sufficient small $\mu_i > 0$, $i=1,\ldots,K$, such that 
$\bI + \sum_{j \neq i} \mu_j \bH_j - \frac{\mu_i}{\gamma_i} \bH_i \succ \bzero$ for all $i$.}
Another result is the following:

\begin{Fact} \label{fac:song_pcsi}
Consider Problem~\eqref{eq:main_pcsi2} with $\sigma_i > 0$ and 
\[ \bH_i = \bh_i \bh_i^H + \bXi_i, \quad \bh_i \neq \bzero, \quad \bXi_i \succeq \bzero, \]
for all $i$.
Suppose that Problem~\eqref{eq:main_pcsi2} is feasible.
If the optimal solution $\bar{\bmu}$ to the dual problem~\eqref{eq:main_pcsi2_dual} satisfies 
\[  1 - \frac{\bar{\mu}_i}{\gamma_i} {\rm Tr}( \bXi_i) > 0, ~i=1,\ldots,K, \]
then the optimal solution $\bar{\cW}$ to Problem~\eqref{eq:main_pcsi2} must be of rank one; i.e., ${\rm rank}(\bar{\bW}_i)= 1$ for all $i$.
\end{Fact}

Fact~\ref{fac:song_pcsi} is essentially a reduction of the robust rank-one result by Song {\em et al.} \cite{Song11} to the fixed-channel case;
see also \cite{Sagnol11,liao2011qos} for similar results that arise in different contexts.
Fact~\ref{fac:song_pcsi} gives a sufficient condition on when the solution to the fixed-channel rate-constrained problem {\em must} have rank one.
In comparison, SDP rank reduction results, such as Fact~\ref{fac:rank_reduce}, usually tell when a rank-one solution to the problem {\em exists} and may not rule out the existence of higher-rank solutions.
Given its importance, we show its proof below.

\medskip
{\em Proof of Fact~\ref{fac:song_pcsi}:} \
Notice that the conditions in Fact~\ref{fac:sd_pcsi} hold.
Thus, the KKT conditions for Problems~\eqref{eq:main_pcsi2} and~\eqref{eq:main_pcsi2_dual}, which are given by
\begin{align}
\sigma_i^2 & \textstyle  \leq {\rm Tr}( \bH_i ( \frac{1}{\gamma_i} \bW_i - \sum_{j \neq i} \bW_j ) ),  
\label{eq:song_pcsi_kkt0} \\
\bm Z_i & = \textstyle \bI + \sum_{j \neq i} \mu_j \bH_j - \frac{\mu_i}{\gamma_i} \bH_i, 
\label{eq:song_pcsi_kkt1} \\
\bW_i & \succeq \bzero, \bm Z_i \succeq \bzero, \mu_i \geq 0, 
\label{eq:song_pcsi_kkt2} \\
\bW_i \bm Z_i & = \bzero, 
\label{eq:song_pcsi_kkt3} \\
0 & = \textstyle \mu_i \left[ \sigma_i^2 - {\rm Tr}( \bH_i ( \frac{1}{\gamma_i} \bW_i - \sum_{j \neq i} \bW_j ) \right], 
\end{align}
for $1=1,\ldots,K$,
are necessary and sufficient for optimality.  For convenience, we use $(\cW, \bmu)$ to denote a primal-dual pair of optimal solutions to Problems~\eqref{eq:main_pcsi2} and~\eqref{eq:main_pcsi2_dual}.
Let us expand 
the $i$th constraint in \eqref{eq:song_pcsi_kkt1} as
\[ 
\bm Z_i =  \textstyle \left(\bI - \frac{\mu_i}{\gamma_i} \bXi_i \right)+ \sum_{j \neq i} \mu_j \bH_j - \frac{\mu_i}{\gamma_i} \bh_i \bh_i^H
\]
for $i=1,\ldots,K$, and recall that $\bH_i \succeq \bzero$.
Observe that if $\bI - \frac{\mu_i}{\gamma_i} \bXi_i \succ \bzero$,
then at least $N-1$ of the eigenvalues of $\bm Z_i$ must be positive.
Consequently, we have ${\rm rank}(\bm Z_i) \geq N-1$.
This, together with \eqref{eq:song_pcsi_kkt2}--\eqref{eq:song_pcsi_kkt3}, imply that ${\rm rank}(\bW_i) \leq 1$.
Also, since $\bW_i= \bzero$ violates \eqref{eq:song_pcsi_kkt0} for $\sigma_i^2 > 0$, we are left with ${\rm rank}(\bW_i) = 1$.
Moreover, the condition for 
$\bI - \frac{\mu_i}{\gamma_i} \bXi_i \succ \bzero$ 
is satisfied if $1 >  \frac{\mu_i}{\gamma_i} {\rm Tr}(\bXi_i)$.
The proof is therefore complete.
\hfill $\blacksquare$
\medskip


The second ingredient is 
an alternative representation of the robust constraints in \eqref{eq:main_b}.
To describe it, consider a generic quadratically constrained quadratic program (QCQP)
\begin{equation} \label{eq:qcqp}
\begin{aligned}
\max_{ \bh \in \Cbb^n } & ~ \bh^H \bA_0 \bh + 2{\rm Re}( \bb_0^H \bh ) + c_0 \\
{\rm s.t.}\, & ~ \bh^H \bA_i \bh + 2{\rm Re}( \bb_i^H \bh ) + c_i  \trianglelefteq_i  0, ~i=1,\ldots,m,
\end{aligned}
\end{equation}
where $\bA_i \in \Hbb^n, \bb_i \in \Cbb^n, c_i \in \Rbb$ for $i=  0,\ldots,m$.
We see that 
the problems in  \eqref{eq:main_b} are instances
of Problem~\eqref{eq:qcqp} with $m=1$.
Problem~\eqref{eq:qcqp} is generally non-convex; e.g., when $\bA_0$ is indefinite (which is the case in \eqref{eq:main_b}).
However, it can be tackled by the SDR technique~\cite{Luo2010_SPM}.
Concisely, SDR works by letting $\bH= \bh \bh^H$, relaxing it to $\bH \succeq \bh \bh^H$, and putting it into \eqref{eq:qcqp} to obtain
\begin{equation} \label{eq:qcqp_sdr}
\begin{aligned}
\max_{ \bH \in \Hbb^n, \bh \in \Cbb^n } & ~ {\rm Tr}( \bA_0 \bH) + 2{\rm Re}( \bb_0^H \bh ) + c_0 \\
{\rm s.t.} & ~  {\rm Tr}( \bA_i \bH) + 2{\rm Re}( \bb_i^H \bh ) + c_i  \trianglelefteq_i 0, ~i=1,\ldots,m, \\
& ~ \bH \succeq \bh \bh^H.
\end{aligned}
\end{equation}
Problem~\eqref{eq:qcqp_sdr} is convex.
Let $\varphi^\star$ and $\phi^\star$ denote the optimal values of Problems~\eqref{eq:qcqp} and \eqref{eq:qcqp_sdr}, respectively. As a relaxation, we 
have
\[ 
\varphi^\star \leq \phi^\star.
\]
However, if $m \leq 2$, then we can have $\varphi^\star = \phi^\star$ under 
some fairly 
mild conditions~\cite{Beck06,Huang07}.
Here we describe one such condition.
\begin{Fact} \label{fac:qcqp2}
Suppose that both Problems~\eqref{eq:qcqp} and \eqref{eq:qcqp_sdr} have optimal solutions.
Then, for $m \leq 2$, the optimal value of Problem~\eqref{eq:qcqp} is equal to that of Problem~\eqref{eq:qcqp_sdr}.
\end{Fact}
Fact~\ref{fac:qcqp2} can be easily deduced by applying the SDP rank reduction result in Fact~\ref{fac:rank_reduce} to Problem~\eqref{eq:qcqp_sdr}; see, e.g., \cite{Huang07} for the proof.

For our problem in \eqref{eq:main_b}, Fact~\ref{fac:qcqp2} holds.
Specifically, it can be verified that the corresponding problems in \eqref{eq:qcqp} and \eqref{eq:qcqp_sdr} have compact feasible sets. Thus, they both have optimal solutions.

%
%

\subsection{Proof of Theorem~\ref{thm:main}: Part One}
\label{sec:part1}

Consider Problem~\eqref{eq:main},
and suppose that it has an optimal solution.
As mentioned in the previous subsection,
we can use the tight SDR result in Fact~\ref{fac:qcqp2} to derive the following equivalent representation of the constraint functions in \eqref{eq:main_b}:
\[
\max_{\bh_i \in \setU_i} \varphi_i(\cW,\bh_i) =  \max_{\bH_i \in \setV_i} \phi_i(\cW,\bH_i),
\quad i=1,\ldots,K,
\]
where 
\ifconfver
    \begin{align}
 &     \phi_i(\cW,\bH_i)  = \textstyle \sigma_i^2 +  {\rm Tr}( \bH_i  (\sum_{j \neq i} \bW_j  - \frac{1}{\gamma_i} \bW_i ) ),
    \nonumber \\
    	 & \setV_i   = \{ \bH_i \in \Hbb^N ~|~ 
	\exists \bh_i \in \Cbb^N \text{~s.t.~} 
	\bH_i \succeq \bh_i \bh_i^H, 
	\nonumber \\
    	  & 
	 \qquad  \qquad 
	\| \bbh_i \|_2^2 - 2 {\rm Re}( \bbh_i^H \bh_i ) + {\rm Tr}(\bH_i) \leq \eps_i^2  \}.
    \label{eq:setV}
    \end{align}
\else
    \begin{align}
    \phi_i(\cW,\bH_i) & = \textstyle \sigma_i^2 +  {\rm Tr}( \bH_i  (\sum_{j \neq i} \bW_j  - \frac{1}{\gamma_i} \bW_i ) ),
    \nonumber \\
    	\setV_i & = \{ \bH_i \in \Hbb^N ~|~ \exists \bh_i \in \Cbb^N \text{~s.t.~} \bH_i \succeq \bh_i \bh_i^H, 
    	~ {\rm Tr}(\bH_i) - 2   {\rm Re}( \bbh_i^H \bh_i ) + \| \bbh_i \|_2^2   \leq \eps_i^2 \}.
    \label{eq:setV}
    \end{align}
\fi
This leads to a new formulation of Problem~\eqref{eq:main} as follows:
\begin{equation} \label{eq:main_sdr}
\begin{aligned}
v^\star= \min_{ \cW \in \setS} & ~ \textstyle \sum_{i=1}^K {\rm Tr}(\bW_i)   \\
{\rm s.t.}\, & ~  \max_{\bH_i \in \setV_i} \phi_i(\cW,\bH_i) \leq 0,  ~ i=1,\ldots,K,  
\end{aligned}
\end{equation}
where
\begin{equation}
\setS  = \{ (\bW_1,\ldots, \bW_K ) \in \Hbb^N \times \cdots \times \Hbb^N ~|~  \bW_i \succeq \bzero 
	~ \forall i \}.
\label{eq:setS}
\end{equation}
We should emphasize a key difference in the new formulation \eqref{eq:main_sdr}:
Every constraint function $\phi_i(\cW,\bH_i)$ in \eqref{eq:main_sdr} is affine in $\bH_i$, given any $\cW \in \setS$. 
In comparison,
the related constraint function $\varphi_i(\cW,\bh_i)$ in \eqref{eq:main} is 
indefinite quadratic
in $\bh_i$, given a general $\cW \in \setS$.

Let $\cW^\star$ denote the optimal solution to \eqref{eq:main_sdr} (throughout the proof we will assume this without further mentioning).
Then, it can be seen from Problem~\eqref{eq:main_sdr} that $\cW^\star$ is a feasible solution to
\begin{equation} \label{eq:rc}
\begin{aligned}
\min_{ \cW \in \setS} & ~ \textstyle \sum_{i=1}^K {\rm Tr}(\bW_i)   \\
{\rm s.t.}\, & ~  \phi_i(\cW,\bH_i) \leq 0,  ~ i=1,\ldots,K
\end{aligned}
\end{equation}
for any given $\cH = ( \bH_1,\ldots,\bH_K ) \in \setV \triangleq  \setV_1 \times \cdots \times \setV_K$. 
Furthermore, given an $\cH \in \setV$, Problem~\eqref{eq:rc} is an instance of the fixed-channel problem in \eqref{eq:main_pcsi2}.
Subsequently, by Fact~\ref{fac:sd_pcsi}, for any given $\cH \in \setV$, Problem~\eqref{eq:rc} has an optimal solution. In particular, we can define
\begin{equation*} 
\begin{aligned}
p(\cH)= \min_{ \cW \in \setS} & ~ \textstyle \sum_{i=1}^K {\rm Tr}(\bW_i)   \\
{\rm s.t.}\, & ~  \phi_i(\cW,\bH_i) \leq 0,  ~ i=1,\ldots,K.
\end{aligned}
\end{equation*}
Since $p(\cH) \leq \sum_{i=1}^K {\rm Tr}(\bW_i^\star) = v^\star$ for any $\cH \in \setV$, we have
\[ v^\star \geq p^\star \triangleq \sup_{ \cH \in \setV } ~ p(\cH). \]
We now develop the following key result:

\begin{Prop} \label{prop:weak_prop}
Suppose that Problem~\eqref{eq:main_sdr} has an optimal solution, and that $\sigma_i^2 > 0$ for all $i$.
The following two properties hold:
\begin{enumerate}
\item The optimal value $p^\star$ is attained; i.e., there exists an $\cH^\star \in \setV$ such that $p^\star= p(\cH^\star)$.
\item Suppose that $v^\star= p^\star$ holds,
and let $\cH^\star$ be an optimal solution to $\sup_{ \cH \in \setV }  p(\cH)$.
Then, the optimal solution $\cW^\star$ to Problem~\eqref{eq:main_sdr} must be an optimal solution to Problem~\eqref{eq:rc} for $\cH = \cH^\star$.
\end{enumerate}
\end{Prop}

\medskip
{\em Proof of Proposition~\ref{prop:weak_prop}:} \
We will show that $p(\cH)$ is upper semicontinuous on $\setV$.
Let us assume this for the time being.
By noting that $\setV$ is compact, we see that $p^\star = \sup_{ \cH \in \setV } ~ p(\cH)$ is attained by some $\cH \in \setV$.  This establishes the first property.
For the second property,
suppose that $v^\star = p^\star$, but $\cW^\star$ is not optimal for Problem~\eqref{eq:rc} when $\cH = \cH^\star$.
Then, we have $p^\star = p(\cH^\star) < \sum_i {\rm Tr}(\bW_i^\star)  = v^\star = p^\star$, a contradiction.

It remains to establish the upper semicontinuity of $p(\cH)$ on $\setV$.
The main tools we need are summarized as follows:

\begin{Fact} \label{fac:lsc}
Let $\mathcal{E},\mathcal{F}$ be finite-dimensional Euclidean spaces.
\begin{enumerate} 
\item[(a)] ( cf. \cite[Proposition 1.20]{Penot13}) Let $\setI$ be an arbitrary index set and $\setX\subset\mathcal{E}$ be arbitrary.  Suppose that $f_i: \setX \rightarrow [-\infty,\infty)$ is upper semicontinuous at $\bar{\bx} \in \setX$ for all $i\in I$.  Then, $f=\inf_{i\in \setI} f_i$ is upper semicontinuous at $\bar{\bx}$.

\item[(b)] (cf. \cite[Corollary 1.23]{Penot13}) Let $\setX\subset\mathcal{E}$ be arbitrary and $\setY\subset\mathcal{F}$ be compact.  Suppose that $f:\setX\times \setY \rightarrow [-\infty,\infty)$ is upper semicontinuous at all points in $\{\bar{\bx}\}\times \setY$.  Then, the function $g:\setX \rightarrow [-\infty,\infty)$ defined by $g(\bx) = \sup_{\by \in \setY} f(\bx,\by)$ is upper semicontinuous at $\bar{\bx}$.
\end{enumerate}
\end{Fact}

Since Fact~\ref{fac:sd_pcsi} implies that Problem~\eqref{eq:rc} attains zero duality gap for any $\cH \in \setV$ (note also the feasibility of Problem~\eqref{eq:rc} for any $\cH \in \setV$ and $\sigma_i^2 > 0$ for all $i$), we can use the dual form in \eqref{eq:main_pcsi2_dual} to equivalently express $p(\cH)$ as
\begin{equation} 
\label{eq:pH_dual}
p(\cH) = \sup_{\bmu \geq \bzero} \inf_{ \cZ \in \setS } ~ g( \cH, \bmu, \cZ ),
\quad \text{for any $\cH \in \setV$,}
\end{equation}
where $\cZ = ( \bZ_1, \ldots, \bZ_K ) \in \Hbb^N \times \cdots \times \Hbb^N$, $\setS$ is given in \eqref{eq:setS}, and 
\ifconfver
\begin{align*}
g( \cH, \bmu, \cZ ) & = \sum_{i=1}^K \sigma_i^2 \mu_i \\
& \quad + \sum_{i=1}^K {\rm Tr}\left( \bZ_i \left( \textstyle \bI + \sum_{j \neq i} \mu_j \bH_j - \frac{\mu_i}{\gamma_i} \bH_i \right) \right).
\end{align*}
\else
\[
g( \cH, \bmu, \cZ ) = \sum_{i=1}^K \sigma_i^2 \mu_i + \sum_{i=1}^K {\rm Tr}\left( \bZ_i \left( \textstyle \bI + \sum_{j \neq i} \mu_j \bH_j - \frac{\mu_i}{\gamma_i} \bH_i \right) \right).
\]
\fi
Additionally, since $p(\cH) \leq v^\star < \infty$ and $\sigma_i^2 > 0$ for all $i$, 
it can be readily verified from \eqref{eq:main_pcsi2_dual} that
any optimal solution $\bmu$ to the outer problem in \eqref{eq:pH_dual} is bounded.
Thus, without loss of generality, we may assume the existence of an $M>0$ such that
\begin{equation*} 
p(\cH) = \sup_{\bzero \leq \bmu \leq M \bone} \inf_{ \cZ \in \setS } ~ g( \cH, \bmu, \cZ ),
\quad \text{for any $\cH \in \setV$,}
\end{equation*}
where $\bone$ denotes an all-one vector.
We are now ready to apply Fact~\ref{fac:lsc}.
Let $\Rbb^n_+$ denote the set of nonnegative numbers in $\Rbb^n$.
Since $g$ is continuous, and hence upper semicontinuous, at any $(\cH,\bmu) \in \setV \times \Rbb^K_+$, 
Fact~\ref{fac:lsc}(a) implies that $\bar{g}(\cH,\bmu) = \inf_{\cZ \in \setS } g(\cH,\bmu,\cZ)$ is upper semicontinuous at any $(\cH,\bmu) \in \setV \times \Rbb^K_+$.
Subsequently, by applying Fact~\ref{fac:lsc}(b), $p(\cH) = \sup_{\bzero \leq \bmu \leq M \bone} ~ \bar{g}( \cH, \bmu )$ is upper semicontinuous at any $\cH \in \setV$.
The proof is therefore complete.
\hfill $\blacksquare$
\medskip

\begin{Prop} \label{prop:strong_dual}
Suppose that Problem~\eqref{eq:main_sdr} has an optimal solution,
and that $\sigma_i^2 > 0$ for all $i$.
It holds that $v^\star = p^\star$.
\end{Prop}

The proof of Proposition~\ref{prop:strong_dual} is similar to that of a duality result in robust optimization \cite[Theorem 4.1]{beck2009duality}\footnote{The result in \cite[Theorem 4.1]{beck2009duality} shows a relation called ``primal worst equals dual best.''
Simply speaking, it shows that for a certain class of robust convex optimization problems, the optimal value is equivalent to that of another problem for optimistic dual maximization.}
and is given as follows:

\medskip
{\em Proof of Proposition~\ref{prop:strong_dual}:} \
Proposition~\ref{prop:strong_dual} is a consequence of the following simplified version of Sion's minimax theorem.

\begin{Fact}[Sion's minimax theorem~\cite{sion1958general,komiya1988elementary}]
Let $\mathcal{X}$, $\mathcal{Y}$ be subsets of a finite-dimensional Euclidean space 
and $f( \cdot, \cdot )$ be a real-valued function on $\mathcal{X} \times \mathcal{Y}$.
If 
\begin{enumerate}
\item $\mathcal{X}$ is convex and compact, $\mathcal{Y}$ is convex,
\item $f(\LargerCdot,\by)$ is lower semicontinuous and 
 convex
for each $\by \in \mathcal{Y}$, and
\item $f(\bx, \LargerCdot)$ is upper semicontinuous and 
 concave
for each $\bx \in \mathcal{X}$,
\end{enumerate}
then $$\min_{ \bx \in  \mathcal{X}} \sup_{ \by \in \mathcal{Y} } f(\bx,\by) = \sup_{ \by \in \mathcal{Y} } \min_{ \bx \in  \mathcal{X}}   f(\bx,\by).$$
\end{Fact}
Now, define
\[ 
f(\cW, \blam, \cH )= \sum_{i=1}^K {\rm Tr}(\bW_i) +  \sum_{i=1}^K \lambda_i \phi_i(\cW,\bH_i),
\]
where $\cW \in \setS$, $\blam \geq \bzero$, and $\cH \in \setV$.
Observe that $f$ is affine (and thus continuous and both convex and concave) either in $\cW$, in $\blam$, or in $\cH$.
Using the above function, the Lagrangian function of Problem~\eqref{eq:main_sdr} can be written as
$$\mathcal{L}( \cW, \blam) = \sup_{ \cH \in \setV } ~ f(\cW, \blam, \cH ).$$
Note that $\mathcal{L}$ is convex lower semicontinuous in $\cW$ and 
affine
in $\blam$, respectively
(cf.  Fact~\ref{fac:lsc}(a) for the lower semicontinuity of $\mathcal{L}$).
Let
$\bar{\setS}= \{ \cW \in \setS ~|~ \sum_i {\rm Tr}(\bW_i) \leq R \}$ for some $R > v^\star$,
and replace 
$\setS$ in Problems~\eqref{eq:main_sdr} and \eqref{eq:rc} by $\bar{\setS}$,
which is without loss of generality.
From the fact that the optimal values of Problems~\eqref{eq:main_sdr} and~\eqref{eq:rc} equal
$$v^\star= \min_{\bW \in \bar{\setS}} \sup_{\blam \geq \bzero}  ~ \mathcal{L}( \cW, \blam),
\quad
p(\cH) = \min_{\bW \in \bar{\setS}} \sup_{\blam \geq \bzero}  ~ f( \cW, \blam,\cH),
$$
respectively,
we have the following chain of equalities:
\begin{subequations}
\begin{align}
v^\star 
& = \sup_{\blam \geq \bzero} \min_{\bW \in \bar{\setS}}  ~ \mathcal{L}( \cW, \blam) \label{eq:sd_1} \\
& = \sup_{\blam \geq \bzero}  \sup_{ \cH \in \setV }  \min_{\bW \in \bar{\setS}} ~ f(\cW, \blam, \cH ) \label{eq:sd_2} \\
& =  \sup_{ \cH \in \setV } \sup_{\blam \geq \bzero} \min_{\bW \in \bar{\setS}} ~ f(\cW, \blam, \cH ) \label{eq:sd_3} \\
& = \sup_{ \cH \in \setV }  \min_{\bW \in \bar{\setS}} \sup_{\blam \geq \bzero}  ~ f(\cW, \blam, \cH ) \label{eq:sd_4} \\
& = \sup_{ \cH \in \setV } ~ p(\cH) = p^\star,
\end{align}
\end{subequations}
where 
\eqref{eq:sd_1}, \eqref{eq:sd_2}, and \eqref{eq:sd_4} are all due to Sion's minimax theorem.\footnote{One can also obtain \eqref{eq:sd_1} and \eqref{eq:sd_4} by strong duality in convex optimization, say, under Slater's condition. However, \eqref{eq:sd_2} requires Sion's result.}
Note that in achieving the above equalities, we have also used the fact that $\bar{\setS}$ is convex compact and $\setV$ is convex.
\hfill $\blacksquare$
\medskip

\subsection{Discussion: A Duality Relationship Revealed in the Part-One Proof}

Before we move to the second part of the proof,
let us discuss a relationship revealed in the proof in the last subsection.
For convenience, we summarize the main points as a theorem.
\begin{Theorem} \label{thm:dual}
Suppose that Problem~\eqref{eq:main} has an optimal solution,
and that $\sigma_i^2 > 0$ for all $i$.
The following equality holds for the equivalent representation of Problem~\eqref{eq:main} in \eqref{eq:main_sdr}:
\begin{equation} \label{eq:sdu}
\begin{aligned}
\min_{ \cW \in \setS} & ~ \textstyle \sum_i {\rm Tr}(\bW_i)   \\
{\rm s.t.}\, & ~  \max_{\bH_i \in \setV_i} \phi_i(\cW,\bH_i) \leq 0,  \\
& ~ \qquad \text{for all $i$}
\end{aligned}
\begin{aligned}
= \max_{\cH \in \setV} \min_{ \cW \in \setS} & ~ \textstyle \sum_i {\rm Tr}(\bW_i)   \\
{\rm s.t.}\, & ~  \phi_i(\cW,\bH_i) \leq 0,  \\
& ~ \qquad \text{for all $i$}
\end{aligned}
\end{equation}
Moreover, an optimal solution $\cW^\star$ 
to the problem
on the left-hand side (LHS) of \eqref{eq:sdu} 
corresponds to an optimal maximin solution to the problem
on the right-hand side (RHS) of \eqref{eq:sdu}; i.e., there exists an $\cH^\star$ such that $(\cH^\star,\cW^\star)$ is a maximin solution to the problem on the RHS of  \eqref{eq:sdu}.
\end{Theorem}
Note that Theorem~\ref{thm:dual} is a consequence of Propositions~\ref{prop:weak_prop} and \ref{prop:strong_dual}.
The equality in \eqref{eq:sdu}
shows strong physical meaning---the robust rate-constrained problem is equivalent, in terms of the optimal value, to a problem where we solve the fixed-channel rate-constrained problems for all (semidefinite-relaxed) channel possibilities $\cH \in \setV$, and then select the one whose minimal total transmit power (w.r.t. $\cW$) is the worst (w.r.t. $\cH$).

It is worth noting that in proving the duality result \eqref{eq:sdu}, we require each channel region $\setV_i$ to be convex compact only. The specific structures of $\setV_i$ have not been used yet.
As we will see, this will give us advantages when we extend our result.

While appealing in implications,
the duality result \eqref{eq:sdu} still does not lead to the desired rank-one result.
Let us discuss this issue by listing the following facts:
\begin{enumerate}
\item 
Given an optimal solution  $\cW^\star$ to the problem on the LHS of \eqref{eq:sdu},
there exists an $\cH^\star$ such that $(\cH^\star,\cW^\star)$ is a maximin solution to the problem on the RHS of  \eqref{eq:sdu}.

\item By SDP rank reduction (e.g., Fact~\ref{fac:rank_reduce}), every inner problem on the RHS of  \eqref{eq:sdu} has a rank-one solution.
Thus, there exists a maximin solution $(\cH',\cW')$ to 
the problem on
the RHS of  \eqref{eq:sdu} such that $\cW'$ has rank-one.
\end{enumerate}

From the above facts, one would be tempted to think that the rank-one result should be within grasp.
Unfortunately, there is a gap.
While the inner problem on the RHS of  \eqref{eq:sdu} admits a rank-one solution $\cW'$,
it may have more than one solution.  Indeed, it is possible that 
a higher-rank $\cW'$ exists and is also a solution to the inner problem.
Consequently, we are unable to tell whether the solution $\cW^\star$ to 
the problem on 
the LHS of \eqref{eq:sdu} is a rank-one solution $\cW'$ to the inner problem on the RHS of \eqref{eq:sdu}.
The only exception is when every  solution to
the inner problem on
 the RHS of \eqref{eq:sdu} is of rank one---that is the direction we will pursue in the second part of the proof.

\medskip
\noindent
{\bf Remark 2.} \
As a more technical question, one may wonder why we use the equivalent SDR representation of the robust problem, Problem~\eqref{eq:main_sdr}, to perform analysis.
Instead, why not use the original problem \eqref{eq:main}, which is more direct?
In fact, 
we can also do that, and
except for one specific point, all the arguments in the last subsection apply.
Let us summarize the result.
\begin{Corollary} \label{cor:wd}
Suppose that Problem \eqref{eq:main} has an optimal solution,
and that $\sigma_i > 0$ for all $i$.
 Then, the following inequality holds:
\begin{equation} \label{eq:wd}
\begin{aligned}
\min_{ \cW \in \setS} & ~ \textstyle \sum_i {\rm Tr}(\bW_i)   \\
{\rm s.t.}\, & ~  \max_{\bh_i \in \setU_i} \varphi_i(\cW,\bh_i) \leq 0,  \\
& ~ \qquad \text{\rm for all $i$}
\end{aligned}
\begin{aligned}
\geq \max_{ \substack{ \bh_i \in \setU_i \\ ~ \forall i}} \min_{ \cW \in \setS} & ~ \textstyle \sum_i {\rm Tr}(\bW_i)   \\
{\rm s.t.}\, & ~  \varphi_i(\cW,\bh_i) \leq 0,  \\
& ~ \qquad \text{\rm for all $i$}
\end{aligned}
\end{equation}
Furthermore, for instances where equality in \eqref{eq:wd} holds,
any optimal solution to Problem \eqref{eq:main} must be of rank one.
\end{Corollary}


{\em Proof of Corollary~\ref{cor:wd}:} \
Since the proof is similar to 
that in the last subsection,
we only describe the key steps.
Let $\bF = [~ \bh_1,\ldots, \bh_K ~]$.
By replacing $p(\cH)$ with
\[
\begin{aligned}
p(\bF)= \min_{ \cW \in \setS} & ~ \textstyle \sum_{i=1}^K {\rm Tr}(\bW_i)   \\
{\rm s.t.}\, & ~  \varphi_i(\cW,\bh_i) \leq 0,  ~ i=1,\ldots,K
\end{aligned}
\]
and $\setV$ with $\setU = \setU_1 \times \cdots \times \setU_K$,
it can be shown that $v^\star \geq \hat{p} \triangleq \sup_{\bF \in \setU } p(\bF)$.
Proposition~\ref{prop:weak_prop} can also be shown to be applicable---i.e., 1)
$\hat{p}$ is attained; 2) if $v^\star = \hat{p}$, then an optimal solution $\cW^\star$ to the problem on the LHS of \eqref{eq:wd} corresponds an optimal maximin solution to the problem on the RHS of \eqref{eq:wd}.
Moreover, by Fact~\ref{fac:song_pcsi}, any optimal solution to the inner problem on the RHS of \eqref{eq:wd} must be of rank one.
Thus, if $v^\star = \hat{p}$, then $\cW^\star$ must be of rank one.
The proof is complete.
\hfill $\blacksquare$
\medskip

Note that in the proof above, the zero duality gap result in
Proposition~\ref{thm:dual} is not applicable. The reason is that $\varphi_i(\cW,\bh_i)$ is non-concave in $\bh_i$, which forbids us from using Sion's minimax theorem at one specific point (cf. \eqref{eq:sd_2}).

Corollary~\ref{cor:wd} leads to a curious question:
Does equality in \eqref{eq:wd} hold for {\em all} problem instances?
A positive answer to this question would imply a strong rank-one result. Unfortunately, we have the following negative result:
\begin{Prop} \label{prop:wd2}
There 
exist
problem instances for which \eqref{eq:wd} holds with strict inequality.
\end{Prop}
The proof of Proposition~\ref{prop:wd2} is 
relegated to Appendix~\ref{apx:wd}. 
Although we disprove the equality in \eqref{eq:wd} in general, Corollary~\ref{cor:wd} may be useful in that it enables one to tackle the rank-one analysis problem by studying conditions under which equality in \eqref{eq:wd} holds.
We leave this as an open direction.

\subsection{Proof of Theorem~\ref{thm:main}: Part Two}
\label{sec:part2}

As discussed above, if we can prove that given any $\cH \in \setV$, the inner problem on the RHS of \eqref{eq:sdu},
recapitulated here as
\begin{equation} \label{eq:main_pcsi3}
\begin{aligned}
\min_{ \cW} & ~ \textstyle \sum_{i=1}^K {\rm Tr}(\bW_i)   \\
{\rm s.t.} 
& ~  \textstyle {\rm Tr}( \bH_i ( \frac{1}{\gamma_i} \bW_i - \sum_{j \neq i} \bW_j ) ) \geq \sigma_i^2,  ~ i=1,\ldots,K,   \\
& ~ \bW_1, \ldots, \bW_K \succeq \bzero,
\end{aligned}
\end{equation}
only admits rank-one solutions,
then any solution to our main problem~\eqref{eq:main} must be of rank one.
To study when this can happen, consider the channel set $\setV_i$ in \eqref{eq:setV}.
Let
\[ \bXi_i = \bH_i - \bh_i \bh_i^H, ~i=1,\ldots,K. \]
By a change of variable, we can equivalently characterize  $\setV_i$ as
\ifconfver
    \begin{align}  \label{eq:setV_2}
     \setV_i & = \{ \bH_i = \bh_i \bh_i^H + \bXi_i \in \Hbb^N ~|~ \bXi_i \succeq \bzero, ~ \bh_i \in \Cbb^N, \nonumber \\
    & \qquad \| \bh_i - \bbh_i \|_2^2 + {\rm Tr}(\bXi_i) \leq \eps_i^2 \}. \nonumber
    \end{align}
\else
    \begin{align} \label{eq:setV_2}
    \setV_i & = \{ \bH_i = \bh_i \bh_i^H + \bXi_i \in \Hbb^N ~|~ \bXi_i \succeq \bzero, 
    ~ \bh_i \in \Cbb^N,
    ~ \| \bh_i - \bbh_i \|_2^2 + {\rm Tr}(\bXi_i) \leq \eps_i^2 \}. \nonumber
    \end{align}
\fi
Let us assume that
\begin{equation} \label{eq:part2_cond1}
\bh_i \neq \bzero, ~i=1,\ldots,K.
\end{equation}
With the above setting, we can apply the rank-one result in Fact~\ref{fac:song_pcsi}:
The solution to Problem~\eqref{eq:main_pcsi3} must be of rank one if 
\begin{equation} \label{eq:part2_cond_suff} 
\frac{\bar{\mu}_k}{\gamma_k} {\rm Tr}(\bXi_k) < 1, ~k=1,\ldots,K,
\end{equation}
where $(\bar{\mu}_1,\ldots,\bar{\mu}_K)$ denotes the optimal solution to the dual of Problem~\eqref{eq:main_pcsi3}, which appears in \eqref{eq:main_pcsi2_dual}.

Since $\bar{\mu}_k$ does not have a closed-form solution,
we wish to prove an analytically tractable bound on $\bar{\mu}_k$. 
To this end, let us denote 
\[ \beta_i = \| \bbPi_i \bbh_i \|_2, \quad \zeta_i = \sqrt{ \eps_i^2 - {\rm Tr}(\bXi_i) } \]
(recall that $\bbPi_i$ is the orthogonal complement projector of $[~ \bbh_1, \ldots, \bbh_{i-1}, \bbh_{i+1},\ldots,\bbh_K ~]$).  We have the following result:

\begin{Prop} \label{prop:dual_bnd}
Consider Problem~\eqref{eq:main_pcsi3}, where $\bH_i \in \setV_i$, $i=1,\ldots,K$.
Let $k \in \{ 1,\ldots,K \}$ be given, and suppose that
\begin{subequations} \label{eq:dual_bnds_cond}
\begin{align}
\beta_i - \zeta_i
& \geq \eps_i 
\sqrt{ \gamma_i (K-1) }, ~ \text{\rm for all $i \neq k$}, \\
\beta_k - \zeta_k
& > \eps_k 
\sqrt{\gamma_k (K-1)}. 
\end{align}
\end{subequations}
Then, any feasible solution $\bmu$ to the dual of Problem~\eqref{eq:main_pcsi3} (or any feasible solution to Problem~\eqref{eq:main_pcsi2_dual}) satisfies 
\begin{equation} \label{eq:dual_bnds_res}
\mu_k \leq \frac{K}{ \frac{1}{\gamma_k} ( \beta_k - \zeta_k)^2 - (K-1)\eps_k^2 }.
\end{equation}
\end{Prop}

\medskip
\noindent
{\em Proof of Proposition~\ref{prop:dual_bnd}:} \
First, consider the problem
\begin{equation} \label{eq:max_muk}
\begin{aligned}
\max_{ \bmu} & ~ \textstyle \mu_k   \\
{\rm s.t.} 
& ~  \textstyle \bI + \sum_{j \neq i} \mu_j \bH_j - \frac{\mu_i}{\gamma_i} \bH_i \succeq \bzero,  ~ i=1,\ldots,K,   \\
& ~ \mu_1, \ldots, \mu_K \geq 0.
\end{aligned}
\end{equation}
Observe that the feasible set of Problem~\eqref{eq:max_muk} is exactly the same as that of Problem~\eqref{eq:main_pcsi2_dual}.
The dual of Problem~\eqref{eq:max_muk} is 
\begin{equation} \label{eq:dual_max_muk}
\begin{aligned}
\min_{ \cW} & ~ \textstyle \sum_{i=1}^K {\rm Tr}(\bW_i)   \\
{\rm s.t.} 
& ~  \textstyle {\rm Tr}( \bH_i ( \frac{1}{\gamma_i} \bW_i - \sum_{j \neq i} \bW_j ) ) \geq 0,  ~ \forall i \neq k,   \\
& ~  \textstyle {\rm Tr}( \bH_k ( \frac{1}{\gamma_k} \bW_k - \sum_{j \neq k} \bW_j ) ) \geq 1, \\
& ~ \bW_1, \ldots, \bW_K \succeq \bzero.
\end{aligned}
\end{equation}
By SDP weak duality,
given any feasible solutions $\bmu$ and $\cW$ to Problems~\eqref{eq:max_muk} and \eqref{eq:dual_max_muk}, respectively, we have 
$\sum_i {\rm Tr}(\bW_i) \geq  \mu_k$.

Second, we construct a feasible solution to Problem~\eqref{eq:dual_max_muk} and use its objective value to bound $\mu_k$.
To be specific, choose
\[ 
 \bW_i = \alpha \bm u_i \bm u_i^H, \quad \bm u_i = \frac{ \bbPi_i \bbh_i }{ \| \bbPi_i \bbh_i \|_2 },  \quad i=1,\ldots,K, \]
where $\alpha > 0$ is to be determined.
Substituting the above $\bW_i$'s into the linear constraints in \eqref{eq:dual_max_muk}, we get
\ifconfver
    \begin{align}
    & \textstyle {\rm Tr}\left( \bH_i \left( \frac{1}{\gamma_i} \bW_i - \sum_{j \neq i} \bW_j \right) \right) \nonumber \\     
    & \textstyle \qquad   = \alpha \cdot \bigg[
       \frac{1}{\gamma_i} \Big( \big|  \| \bar{\bPi}_i \bbh_i \|_2 + \bm u_i^H \be_i \big|^2 + \bm u_i^H \bXi_i \bm u_i \Big) \nonumber \\
     & \textstyle \qquad\qquad  
     - \sum_{j \neq i} ( | \bm u_j^H \be_i |^2 + \bm u_j^H \bXi_i \bm u_j ) \bigg],
     \label{eq:r1_3.2} 
     \end{align}
\else
    \begin{align}
    {\rm Tr}\left( \bH_i \left( \frac{1}{\gamma_i} \bW_i - \sum_{j \neq i} \bW_j \right) \right) 
       & = \alpha \cdot \left[
       \frac{1}{\gamma_i} \Big( \big|  \| \bar{\bPi}_i \bbh_i \|_2 + \bm u_i^H \be_i \big|^2 + \bm u_i^H \bXi_i \bm u_i \Big) - \sum_{j \neq i} ( | \bm u_j^H \be_i |^2 + \bm u_j^H \bXi_i \bm u_j ) \right],
       \label{eq:r1_3.2}
    \end{align}
 \fi
where $\be_i = \bh_i - \bbh_i$.
 By noting that $| \bm u_j^H \be_i |^2 \leq \| \bm u_j \|_2^2 \| \be_i \|_2^2 \leq \eps_i^2 - {\rm Tr}(\bXi_i) =  \zeta_i^2$
 and assuming $\beta_i =  \| \bar{\bPi}_i \bbh_i \|_2 \geq \zeta_i$, which is satisfied under \eqref{eq:dual_bnds_cond}, we have
 \begin{align*}
 |  \| \bar{\bPi}_i \bbh_i \|_2 + \bm u_i^H \be_i | & \geq \beta_i - | \bm u_i^H \be_i | \geq \beta_i - \zeta_i, \\
 | \bm u_j^H \be_i |^2 + \bm u_j^H \bXi_i \bm u_j & \leq \zeta_i^2 + {\rm Tr}(\bXi_i) = \eps_i^2.
 \end{align*}
By plugging the above inequalities and the inequality $\bm u_i^H \bXi_i \bm u_i \geq 0$ into \eqref{eq:r1_3.2}, we further get
 \ifconfver
    \begin{align*}
    & \textstyle {\rm Tr}\left( \bH_i \left( \frac{1}{\gamma_i} \bW_i - \sum_{j \neq i} \bW_j \right) \right)  \\    
    & \textstyle \qquad  
     \geq \alpha \cdot \left[
       \frac{1}{\gamma_i} ( \beta_i - \zeta_i )^2 - (K-1) \eps_i^2 \right].
     \end{align*}
\else
    \begin{align*}
    {\rm Tr}\left( \bH_i \left( \frac{1}{\gamma_i} \bW_i - \sum_{j \neq i} \bW_j \right) \right) 
       & \geq \alpha \cdot \left[
       \frac{1}{\gamma_i} ( \beta_i - \zeta_i )^2 - (K-1) \eps_i^2 \right].
    \end{align*}
 \fi
From the above inequality and by the assumptions in \eqref{eq:dual_bnds_cond}, 
one can verify that $\cW$ is feasible for Problem \eqref{eq:dual_max_muk} if 
\[ \alpha = 1 / \left( \textstyle \frac{1}{\gamma_k}  (\beta_k - \zeta_k)^2 - (K-1) \eps_k^2 \right). \]
 Finally, by observing that $\sum_i {\rm Tr}( \bW_i ) = \alpha K$, we obtain \eqref{eq:dual_bnds_res}.
The proof is complete.
\hfill $\blacksquare$
\medskip

Let us return to the sufficient condition in \eqref{eq:part2_cond_suff}.
Assume that \eqref{eq:dual_bnds_cond} holds for all $k$.
By Proposition~\ref{prop:dual_bnd}, the LHS of \eqref{eq:part2_cond_suff} is bounded by
\[ \frac{\bar{\mu}_k}{\gamma_k} {\rm Tr}(\bXi_k) \leq 
\frac{K( \eps_k^2 - \zeta_k^2 )}{ ( \beta_k - \zeta_k )^2 - (K-1) \gamma_k \eps_k^2}.
\]
It follows that if 
\begin{equation} \label{eq:final_bnd_c1}
\frac{K( \eps_k^2 - \zeta_k^2 )}{ ( \beta_k - \zeta_k )^2 - (K-1)\gamma_k \eps_k^2} < 1 
\end{equation}
for all $0\leq \zeta_k \leq \eps_k$ and $k=1,\ldots,K$, then \eqref{eq:part2_cond_suff} will be satisfied for all $\cH \in \setV$ and the desired rank-one result will be achieved.
Before we further analyze \eqref{eq:final_bnd_c1}, we should mention that \eqref{eq:final_bnd_c1} implies  \eqref{eq:part2_cond1} and \eqref{eq:dual_bnds_cond}.  Hence, we no longer require \eqref{eq:part2_cond1}  and \eqref{eq:dual_bnds_cond}  as far as satisfying \eqref{eq:final_bnd_c1} is concerned.
The condition in \eqref{eq:final_bnd_c1} can be reorganized as 
\begin{equation} \label{eq:final_bnd_c2}
\min_{0 \leq \zeta_k \leq \eps_k} \left[ ( \beta_k - \zeta_k )^2 + K \zeta_k^2 \right] > K \eps_k^2 + (K-1) \gamma_k \eps_k^2.
\end{equation}
It is easy to show that
\begin{align*}
\min_{0 \leq \zeta_k \leq \eps_k} \left[ ( \beta_k - \zeta_k )^2 + K \zeta_k^2 \right]
   & \geq \min_{\zeta_k \in \Rbb} \left[ ( \beta_k - \zeta_k )^2 + K \zeta_k^2 \right]  \\
   & = \frac{\beta_k^2 K }{K+1}.
\end{align*}
From the above,  we see that \eqref{eq:final_bnd_c2} is satisfied if 
\begin{equation} \label{eq:final_bnd_c3}
\beta_k^2 \geq \frac{K+1}{K} ( K \eps_k^2 + (K-1) \gamma_k \eps_k^2 ).
\end{equation}
The above inequality is condition \eqref{eq:main_cond} in Theorem~\ref{thm:main}. Thus, we have completed the proof of Theorem~\ref{thm:main}.
We should also mention a refined version of \eqref{eq:final_bnd_c2}.
It can be verified that
\ifconfver
    \begin{align*}
    & \min_{0 \leq \zeta_k \leq \eps_k} \left[ ( \beta_k - \zeta_k )^2 + K \zeta_k^2 \right]  \\
    & \qquad \qquad =
    \left\{
    \begin{array}{ll}
    ( \beta_k - \eps_k )^2 + K \eps_k^2, & \frac{\beta_k}{\eps_k} > K+1, \\
    \frac{\beta_k^2 K }{K+1}, & \frac{\beta_k}{\eps_k} \leq K+1.
    \end{array}
    \right.
    \end{align*}
\else
    \[
    \min_{0 \leq \zeta_k \leq \eps_k} \left[ ( \beta_k - \zeta_k )^2 + K \zeta_k^2 \right] =
    \left\{
    \begin{array}{ll}
    ( \beta_k - \eps_k )^2 + K \eps_k^2, & \frac{\beta_k}{\eps_k} > K+1, \\
    \frac{\beta_k^2 K }{K+1}, & \frac{\beta_k}{\eps_k} \leq K+1.
    \end{array}
    \right.
    \]
\fi
Hence, if $\beta_k/\eps_k > K+1$, then we can replace \eqref{eq:final_bnd_c3} by 
\[ 
( \beta_k - \eps_k )^2 \geq (K - 1) 
\gamma_k 
\eps_k^2.
\]
The above condition leads to \eqref{eq:main_cond_var1} in Remark 1.

\section{Application to Other Channel Error Models}
\label{sec:ext}

Our study in the previous sections has focused on the robust rate-constrained problem under spherically bounded channel errors.
In this section we discuss how the main result can be applied to some other channel error models.


\subsection{The Ellipsoidally Bounded Model}

As a variant of the spherically bounded model, one can also consider the 
ellipsoidally bounded
model~\cite{ZhengWongNg_2008,Zheng_etal2009}
\begin{equation} \label{eq:ellip_U}
\setU_i = \{ \bh_i \in \Cbb^N ~|~ \| \bC^{-\frac{1}{2}}_i (\bh_i - \bbh_i) \|_2 \leq 1 \},
\end{equation}
where $\bC_i \in \Hbb^N$ is given and positive definite,
and $\bC^{\frac{1}{2}}_i$ denotes the positive semidefinite square root of $\bC_i$.
The 
ellipsoidally bounded
model is useful when the base station has some prior knowledge of how the channel errors are spread in the correlation sense.
Note that the 
ellipsoidally bounded
model reduces to the spherically bounded model when $\bC_i = \eps_i^2 \bI$,
and that the eigenvalues of $\bC_i$ are the semi-axis lengths of the ellipsoidal region $\setU_i$.
It has been shown that the corresponding robust rate-constrained problem can also be reformulated as an SDP by using the $\mathcal{S}$-lemma; cf.~\eqref{eq:main_sdp} and see \cite{ZhengWongNg_2008,Zheng_etal2009} for details.  Moreover, we have the following result:
\begin{Corollary} \label{cor:ellip}
Consider Problem~\eqref{eq:main} under the channel sets in \eqref{eq:ellip_U}.
The rank-one result in Theorem~\ref{thm:main} holds if \eqref{eq:main_cond} is replaced by
\begin{equation} \label{eq:main_cond_ellip}
\frac{ \| \bbPi_k \bbh_k \|_2^2 }{ \lambda_{\rm max}(\bC_k) } > 1+K + \left( K - \frac{1}{K} \right) \gamma_k,  \quad k=1,\ldots,K,
\end{equation}
where $\lambda_{\rm max}(\bC_k)$ denotes the largest eigenvalue of $\bC_k$.
\end{Corollary}
The above corollary suggests that the impact of the 
ellipsoidally bounded
model on the rank-one condition lies in the largest semi-axis length $\lambda_{\rm max}(\bC_i)$ of the ellipsoids. 

\medskip
\noindent
{\em Proof of Corollary~\ref{cor:ellip}:} \
The part-one proof of Theorem~\ref{thm:main} in Section~\ref{sec:part1} directly applies.
The reason is that the semidefinite-relaxed channel sets $\setV_i$ of $\setU_i$,
given in this case as
\ifconfver
    \begin{align*}  
     \setV_i & = \{ \bH_i = \bh_i \bh_i^H + \bXi_i  ~|~ \bXi_i \succeq \bzero, ~ \bh_i \in \Cbb^N, \nonumber \\
    & \qquad \| \bC_i^{-\frac{1}{2}} ( \bh_i - \bbh_i ) \|_2^2 + {\rm Tr}( \bC_i^{-\frac{1}{2}} \bXi_i \bC_i^{-\frac{1}{2}} ) \leq 1 \},  
    \end{align*}
\else
    \begin{align*} 
    \setV_i & = \{ \bH_i = \bh_i \bh_i^H + \bXi_i  ~|~ \bXi_i \succeq \bzero, 
    ~ \bh_i \in \Cbb^N,
    ~ \| \bC_i^{-\frac{1}{2}} ( \bh_i - \bbh_i ) \|_2^2 + {\rm Tr}( \bC_i^{-\frac{1}{2}} \bXi_i \bC_i^{-\frac{1}{2}} ) \leq 1 \},
    \end{align*}
\fi
are only required to be convex and compact in the part-one proof.
For the part-two proof in Section~\ref{sec:part2},
recall that Problem~\eqref{eq:main} is guaranteed to admit a rank-one solution if
the optimal solution to the fixed-channel rate-constrained problem in \eqref{eq:main_pcsi3} must be of rank one for any $\cH \in \setV$.
Let
\ifconfver
    \begin{align*}  
     \bar{\setV}_i & = \{ \bH_i = \bh_i \bh_i^H + \bXi_i  ~|~ \bXi_i \succeq \bzero, ~ \bh_i \in \Cbb^N, \nonumber \\
    & \qquad \|  \bh_i - \bbh_i  \|_2^2 + {\rm Tr}(  \bXi_i  ) \leq \lambda_{\rm max}(\bC_i) \}.  
    \end{align*}
\else
    \begin{align*} 
    \bar{\setV}_i & = \{ \bH_i = \bh_i \bh_i^H + \bXi_i  ~|~ \bXi_i \succeq \bzero, 
    ~ \bh_i \in \Cbb^N,
    ~ \| \bh_i - \bbh_i  \|_2^2 + {\rm Tr}( \bXi_i  ) \leq \lambda_{\rm max}(\bC_i) \}.
    \end{align*}
\fi
It is easy to see that $\setV_i \subseteq \bar{\setV}_i$.
Furthermore, from the part-two proof, it is immediate that \eqref{eq:main_cond_ellip} implies that
the optimal solution to
Problem~\eqref{eq:main_pcsi3}  must be of rank one for any $\cH \in \bar{\setV}_1 \times \cdots \times \bar{\setV}_K$.
Consequently, \eqref{eq:main_cond_ellip} also implies the same rank-one result for any $\cH \in \setV$,
and the proof is complete.
\hfill $\blacksquare$

\subsection{A Modified Spherically Bounded Model for FDD}

Consider a specialized model for limited channel feedback in the FDD system~\cite{medra2016low}.
In this context, each user $i$ is pre-assigned a channel direction codebook $\setC_i = \{ \bv_{i,1}, \ldots, \bv_{i,L} \}$, where every codebook element $\bv_{i,l}$ satisfies $\| \bv_{i,l} \|_2 = 1$ and $L$ is the codebook size.
The user estimates the channel $\bh_i$
and feeds back two quantities to the base station,
namely, the channel norm 
$\| \bh_i \|_2$
and the codebook-quantized channel direction
$\hat{\bv}_i = \arg \max_{ \bv \in \setC_i } | \bh_i^H \bv |/ \| \bh_i \|_2$.
Consequently, the presumed channel is 
$\bbh_i = \| \bh_i \|_2 \cdot \hat{\bv}_i$.
If we assume that the channel direction quantization is the dominant source of error, then
we may model the channel error $\be_i = \bh_i - \bbh_i$ as
\[
\| \bbh_i + \be_i \|_2 = \|  \bbh_i  \|_2, 
\quad \frac{ \| \be_i \|_2 }{ \| \bbh_i \|_2 } \leq \delta,
\]
where $\delta > 0$ describes a bound on the channel direction quantization error.
The corresponding channel set $\setU_i$ is
\begin{equation} \label{eq:FDD_U}
\setU_i = \{ \bh_i \in \Cbb^N ~|~ \| \bh_i - \bbh_i \|_2 \leq \delta \| \bbh_i \|_2, ~ \| \bh_i \|_2 = \| \bbh_i \|_2 \}.
\end{equation}
For the above model, an extended form of the $\mathcal{S}$-lemma can be established to deal with the corresponding robust rate-constrained problem~\cite{medra2016low}.
With the extended $\mathcal{S}$-lemma, we have the same development as in Section~\ref{sec:RRC}.
Again, our interest here lies in the rank-one condition.

\begin{Corollary} \label{cor:FDD}
Consider Problem~\eqref{eq:main} under the channel sets in \eqref{eq:FDD_U}.
The rank-one result in Theorem~\ref{thm:main} holds if \eqref{eq:main_cond} is replaced by
\begin{equation} \label{eq:main_cond_FDD}
\frac{ \| \hat{\bPi}_k \hat{\bh}_k \|_2^2 }{ \delta^2 } > 1+K + \left( K - \frac{1}{K} \right) \gamma_k,  \quad k=1,\ldots,K,
\end{equation}
where $\hat{\bh}_i = \bbh_i/\| \bbh_i \|_2$, and $\hat{\bPi}_i$ is the orthogonal complement projector of $[~ \hat{\bh}_1, \ldots, \hat{\bh}_{i-1}, \hat{\bh}_{i+1},\ldots,\hat{\bh}_K ~]$.
\end{Corollary}

We should note that 
$\hat{\bh}_i$ in the above corollary is the channel direction, 
and following the aforementioned system model $\hat{\bh}_i$ lies in the codebook; viz.
$\hat{\bh}_i = \hat{\bv}_i \in \setC_i$.
Corollary~\ref{cor:FDD} suggests that if the chosen codebook elements $\hat{\bv}_i$ among different users are not too similar, and if the codebook resolution is high so that $\delta$ is small,
then the rank-one condition can be achieved.
Also, unlike the result for the spherically bounded model,  the rank-one condition in \eqref{eq:main_cond_FDD} does not depend on the channel magnitude $\| \bbh_i \|_2^2$.

The proof of Corollary~\ref{cor:FDD} is omitted for brevity, since 
it
is similar to, and in fact fundamentally no different from, the proof of Theorem~\ref{thm:main}.
Simply speaking, 
we utilize the fact that 
the channel set $\setU_i$ in \eqref{eq:FDD_U} is spherically bounded 
and apply the same argument as in our previous proof to obtain the rank-one result.
In the proof, the only point that needs some attention
is to show that the SDR representation of the robust constraints in Problem~\eqref{eq:main_sdr} is tight; i.e.,
\[
\max_{ \bh_i \in \setU_i} \phi_i(\cW,\bh_i) = \max_{ \bH_i \in \setV_i} \varphi_i(\cW,\bH_i),
\]
where $\setV_i$ is the semidefinite-relaxed counterpart of $\setU_i$ in \eqref{eq:FDD_U}.
Since $\setU_i$ in \eqref{eq:FDD_U} is defined by two quadratic constraints, it can be verified using Fact~\ref{fac:qcqp2} that the above equality holds.

\subsection{The Box Bounded Model}

Now, let us consider the 
box bounded
model
\begin{equation} \label{eq:box_U}
\setU_i = \{ \bh_i \in \Cbb^N ~|~ | \| \bh_i - \bbh_i \|_\infty \leq \delta_i \},
\end{equation}
where 
the channel error is modeled as being elementwise-bounded by some given $\delta_i > 0$.
This model may be useful when the channel is scalar-quantized and fed back to the base station.
In this case the SDR representation of the robust constraints is no longer tight; i.e., we only have
\[ 
\max_{ \bh_i \in \setU_i } \varphi_i(\cW,\bh_i) \leq \max_{\bH_i \in \setV_i} \phi_i(\cW,\bH_i),
\]
where $\setV_i =  \{ \bH_i \in \Hbb^N ~|~ \bH_i \succeq \bh_i \bh_i^H, ~ \bh_i \in \Cbb^N,
	~ [ \bH_i ]_{jj} - 2  {\rm Re}( [\bbh_i]_j^* [\bh_i]_j )
	+ | [ \bbh_i ]_j |^2   \leq \delta_i^2, ~j=1,\ldots,N \}$,
in general. However, Problem~\eqref{eq:main_sdr}, recapitulated here as
  \begin{subequations} \label{eq:main_sdr_box}
\begin{align}
\min_{ \cW \in \setS} & ~ \textstyle \sum_{i=1}^K {\rm Tr}(\bW_i)   \\
{\rm s.t.} & ~  \max_{\bH_i \in \setV_i} \phi_i(\cW,\bH_i) \leq 0,  ~ i=1,\ldots,K,  \label{eq:main_sdr_box_b}
\end{align}
\end{subequations}
is still useful as a safe approximation; i.e., any feasible solution to Problem~\eqref{eq:main_sdr_box} is also feasible for the original problem~\eqref{eq:main}.
Like the previous cases, Problem~\eqref{eq:main_sdr_box} can be efficiently solved.
By considering the dual of the problem on the LHS of \eqref{eq:main_sdr_box_b} and exploiting zero duality gap,
it can be shown that the constraints in \eqref{eq:main_sdr_box_b} are equivalent to 
\begin{equation*}
\bm t_i \geq \bzero, 
~
\begin{bmatrix}
\bQ_i + {\rm Diag}(\bm t_i)  & \br_i \\
\br_i^H & \ s_i - \delta_i^2 \bm 1^T \bm t_i 
\end{bmatrix} \succeq \bzero,
~ i=1,\ldots,K,
\end{equation*}
where, as before, we have $\bQ_i = \textstyle \frac{1}{\gamma_i} \bW_i - \sum_{j \neq i} \bW_j$,
$\br_i = \bQ_i \bbh_i$,  $s_i = \bbh_i^H  \bQ_i \bbh_i - \sigma_i^2$.
Hence, Problem~\eqref{eq:main_sdr_box} can be rewritten as an SDP, and once again, its solution can be efficiently computed via conic optimization software.

The rank-one result for the box bounded model is as follows:

\begin{Corollary} \label{cor:box}
Consider Problem~\eqref{eq:main_sdr_box}, 
which is
a safe approximation of Problem~\eqref{eq:main} under the channel sets in \eqref{eq:box_U}.
The rank-one result in Theorem~\ref{thm:main} holds if \eqref{eq:main_cond} is replaced by
\begin{equation*} \label{eq:main_cond_box}
\frac{ \| \bbPi_k \bbh_k \|_2^2 }{  N \delta^2 } > 1+K + \left( K - \frac{1}{K} \right) \gamma_k,  \quad k=1,\ldots,K.
\end{equation*}
\end{Corollary}
Again, we shall omit the proof of Corollary~\ref{cor:box} for brevity.
The intuitive idea behind the proof is nothing more than applying
the implication
$\| \bh_i - \bbh_i \|_\infty \leq \delta_i  \Longrightarrow \| \bh_i - \bbh_i \|_2 \leq \sqrt{N} \delta_i$.

\section{Conclusion}
\label{sec:con}

In this paper we provided a rank-one solution analysis for a robust multiuser MISO transmit optimization problem.
Our result suggested that under some practically reasonable condition, the robust problem is guaranteed to admit a rank-one solution.
Our analysis is based on a novel duality framework developed in this paper.
The duality result reveals that the robust problem has a strong connection to another problem that takes a maximin form,
and through that connection we identified a sufficient condition under which the robust problem must have rank-one solutions.
We also discussed how the duality result can be applied to several other robust problems that use different channel error models.
As a future direction, it would be interesting to investigate how the duality result can be extended to deal with a wider class of robust transmit optimization problems, such as the outage-based robust problems~\cite{wang2014outage} and the multicell problems~\cite{shen2012distributed}.

\appendix

\ifplainver
    \section*{Appendix}
    \renewcommand{\thesubsection}{\Alph{subsection}}
\else
    \section{Appendix}
\fi

\subsection{The SDP Rank Reduction Result for Problem~\eqref{eq:main_sdp}}
\label{apx:sdp_rank_reduce}

The aim of this appendix is to apply the SDP rank reduction result in Fact~\ref{fac:rank_reduce} to Problem \eqref{eq:main_sdp} to obtain the solution rank result in \eqref{eq:rank_bnd_robust2}.
The derivations are divided into three steps.

\medskip
{\em Step 1:} \ We reformulate Problem~\eqref{eq:main_sdp} as Problem \eqref{eq:sdp_gen}, and thereby use 
Fact~\ref{fac:rank_reduce}
to deduce a rank result.
Observe that 
each constraint in \eqref{eq:main_sdp_lmi} can be represented by
\begin{subequations} \label{eq:ReIm_z}
\begin{align}
{\rm Re}( [ \bZ_i ]_{k,l} )  & =  \sum_{j,p,q} \left( a_{j,p,q}^{(i,k,l)} {\rm Re}( [ \bW_j ]_{p,q}) + b_{j,p,q}^{(i,k,l)} {\rm Im}([ \bW_j ]_{p,q}) \right) + c^{(i,k,l)} t_i + d^{(i,k,l)}, 
\quad 
\text{for all $k \leq l$,}
\label{eq:Re_z}
\\
{\rm Im}( [ \bZ_i ]_{k,l} )  & =  \sum_{j,p,q} \left( \bar{a}_{j,p,q}^{(i,k,l)} {\rm Re}( [ \bW_j ]_{p,q}) + \bar{b}_{j,p,q}^{(i,k,l)} {\rm Im}([ \bW_j ]_{p,q}) \right) + \bar{c}^{(i,k,l)} t_i + \bar{d}^{(i,k,l)}, 
\quad 
\text{for all $k < l$,}
\label{eq:Im_z}
\end{align}
\end{subequations}
and for some coefficients $a_{j,p,q}^{(i,k,l)}$,  $b_{j,p,q}^{(i,k,l)}$, $c^{(i,k,l)}$, $d^{(i,k,l)}$, $\bar{a}_{j,p,q}^{(i,k,l)}$, $\bar{b}_{j,p,q}^{(i,k,l)}$, $\bar{c}^{(i,k,l)}$, $\bar{d}^{(i,k,l)}$.
Note that there are totally $(N+1)^2$ equations in \eqref{eq:ReIm_z}.
Let us  denote 
\[
\begin{blockarray}{*{7}{c}}
~ & ~ & k & ~&  l & \\
~ & ~ & \downarrow & ~ &  \downarrow & \\
\begin{block}{c[*{5}{c}]c}
\\
~ & ~ & ~ & ~ & \frac{1}{2} & ~ & \leftarrow k  \\
\bF_{k,l}= & \\
~ & ~ & \frac{1}{2} & ~ & ~ & ~ & \leftarrow l  \\
\\
\end{blockarray} \qquad
\begin{blockarray}{*{7}{c}}
~ & ~ & k & ~&  l & \\
~ & ~ & \downarrow & ~ &  \downarrow & \\
\begin{block}{c[*{5}{c}]r}
\\
~ & ~ & ~ & ~ & - \frac{\bm j}{2} & ~ & \leftarrow k  \\
\bG_{k,l}= &  \\
~ & ~ & \frac{\bm j}{2} & ~ & ~ & ~ & \leftarrow l  \\
\\
\end{blockarray}
\]
where the empty entries are all zeros.
It can be verified that given a Hermitian matrix $\bX$,
we have the identities
\[
{\rm Re}( X_{k,l} ) = {\rm Tr}( \bF_{k,l} \bX), \qquad 
{\rm Im}( X_{k,l} ) = {\rm Tr}( \bG_{k,l} \bX). 
\]
Using the above identities, the equations in \eqref{eq:ReIm_z} can be re-expressed as
\begin{subequations} \label{eq:ReIm_z2}
\begin{align}
{\rm Tr}( \bF_{k,l} \bZ_i )
& = \sum_j {\rm Tr}\left( \left( \sum_{p,q} a_{j,p,q}^{(i,k,l)} \bF_{k,l} +  b_{j,p,q}^{(i,k,l)} \bG_{k,l} \right)  \bW_j \right)
+ c^{(i,k,l)} t_i + d^{(i,k,l)},
\quad
\text{for all $k \leq l$,}
\\
{\rm Tr}( \bG_{k,l} \bZ_i )
& = \sum_j {\rm Tr}\left( \left( \sum_{p,q} \bar{a}_{j,p,q}^{(i,k,l)} \bF_{k,l} +  \bar{b}_{j,p,q}^{(i,k,l)} \bG_{k,l} \right)  \bW_j \right)
+ \bar{c}^{(i,k,l)} t_i + \bar{d}^{(i,k,l)},
\quad
\text{for all $k < l$.}
\end{align}
\end{subequations}
We see that every equation in \eqref{eq:ReIm_z2} takes the form in \eqref{eq:sdp_gen_con}.

Moreover, we should note that Fact~\ref{fac:rank_reduce} can be extended to handle situations where the sizes of $\bX_i$'s are unequal; i.e., $\bX_i \in \Hbb^{n_i}$, where $n_i > 0$ can be unequal w.r.t. $i$.
As alluded to in the SDP rank reduction proof, e.g., that of \cite{Huang10TSP,lemon2016low}, such an extension is almost immediate.
Now, consider connecting Problem~\eqref{eq:main_sdp} and Problem~\eqref{eq:sdp_gen} via setting $m= K(N+1)^2$,  $\bX_i= \bW_i, \bX_{i+K}= \bZ_i, \bX_{i+2K}= t_i$ for $i=1,\ldots,K$,
and $k= 3K$.
We see that Problem~\eqref{eq:main_sdp} is equivalent to Problem~\eqref{eq:sdp_gen}.
Hence, by  Fact~\ref{fac:rank_reduce}, we have the following rank result:
\begin{equation} \label{eq:rank_bnd_robust}
\sum_{i=1}^K {\rm rank}(\bW_i^\star)^2 + \sum_{i=1}^K {\rm rank}(\bZ_i^\star)^2 + \sum_{i=1}^K {\rm rank}(t_i^\star)^2 \leq K(N+1)^2.
\end{equation}

\medskip
{\em Step 2:} \
We show that \eqref{eq:rank_bnd_robust} can be reduced to \eqref{eq:rank_bnd_robust2}.
The idea is to prove $t^\star_i > 0$ for all $i$, which, when applied to \eqref{eq:rank_bnd_robust}, results in  \eqref{eq:rank_bnd_robust2}. 
The proof for $t_i^\star > 0$ is as follows.
Suppose that $t_i^\star = 0$.
Let us simply denote $\bQ_i = \textstyle \frac{1}{\gamma_i} \bW_i^\star - \sum_{j \neq i} \bW_j^\star$,
$\br_i = \bQ_i \bbh_i$,  $s_i = \bbh_i^H  \bQ_i \bbh_i - \sigma_i^2$.
By applying the Schur complement to \eqref{eq:main_sdp_lmi}, with $t_i^\star = 0$, we get
$s_i - \br_i^H \bQ_i^\dag \br_i \geq 0$. 
On the other hand, we have
\[ s_i - \br_i^H \bQ_i^\dag \br_i = \bbh_i^H  \bQ_i \bbh_i - \sigma_i^2 - \bbh_i^H \bQ_i \bbh_i = - \sigma_i^2 < 0. \]
Thus, by contradiction, we must not have $t_i^\star = 0$.

\medskip
{\em Step 3:} \
We complete the proof by showing  ${\rm rank}(\bZ_i^\star) \leq N$.
Recall that the size of $\bZ_i$ is $(N+1) \times (N+1)$.
Suppose that ${\rm rank}(\bZ_j^\star) = N+1$ for some $j$.
Then, we can show that there exists a feasible solution that yields a lower objective value than that of $\cW^\star$, a contradiction.
To prove this, assume $j=1$ for convenience.
Let $\bW_1' = (1 - \alpha) \bW_1^\star$ for some $0 < \alpha < 1$, $\bW_i' = \bW_i^\star$, $i=2,\ldots,K$,
and $\bm t' = \bm t^\star$.
The corresponding $\bZ_i$'s in \eqref{eq:main_sdp_lmi} are given by
\[
\bZ_i' = \left\{ \begin{array}{ll} 
\bm Z^\star_1 - \frac{\alpha}{\gamma_1} \bm B_1, & i = 1, \\
\bm Z^\star_i  + \alpha \bm B_i, & \text{otherwise},
\end{array}
\right.
\]
where 
\[
\bm B_i = 
\begin{bmatrix}
\bW_1^\star & \bW_1^\star \bbh_i \\ \bbh_i^H \bW_1^\star  & \bbh_i^H  \bW_1^\star \bbh_i
\end{bmatrix} \succeq \bzero.
\]
Since $\bZ_1^\star$ has full rank, or is positive definite, there exists a sufficiently small $\alpha$ such that $\bm Z^\star_1 - \frac{\alpha}{\gamma_1} \bm B_1 \succeq \bzero$ is satisfied.
Also, $\bm Z^\star_i  + \alpha \bm B_i \succeq \bzero$ holds by nature.
Therefore, for a sufficiently small $\alpha$, $(\bW_i', \bZ_i', t_i')_{i=1}^K$ is a feasible solution to Problem~\eqref{eq:main_sdp}.
Since $\sum_i {\rm Tr}(\bW_i') < \sum_i {\rm Tr}(\bW_i^\star)$ (for $\bW_1^\star \neq \bzero$, which can be easily verified), 
we obtain the desired result.

\subsection{Proof of \eqref{eq:main_imp0}}
\label{apx:eq:main_imp0}

Let $\hat{\bh}_i= \bbh_i / \| \bbh_i \|_2$. We have
\begin{align*}
\| \bbPi_k \bbh_k \|_2 & = \| \bbh_k \|_2 \cdot \| \bbPi_k \hat{\bh}_k \|_2 \\
& = \| \bbh_k \|_2 \cdot \min_{ \bx \in \Cbb^{K-1} } \| \bbF_{-k} \bx - \hat{\bh}_k \|_2 \\
& \equiv \| \bbh_k \|_2 \cdot \min_{ \by \in \Cbb^{K}, ~ y_K = -1}  \| \hat{\bF} \by \|_2 \\
& \geq  \| \bbh_k \|_2 \cdot \min_{ \| \by \|_2 = 1}  \| \hat{\bF} \by \|_2.
\end{align*}
Since
$\min_{ \| \by \|_2 = 1}  \| \hat{\bF} \by \|_2 = \sigma_{\rm min}(\hat{\bF})$ for tall or square $\hat{\bF}$, 
we obtain \eqref{eq:main_imp0}.

\subsection{Proof of Proposition~\ref{prop:prob}}
\label{app:prop:prob}

Let $\setE_k$ denote the event that the $k$th inequality of \eqref{eq:main_cond} is violated; i.e.,
\[
\frac{ \| \bbPi_k \bbh_k \|_2^2 }{ \eps_k^2 } \leq 1+K + \left( K - \frac{1}{K} \right) \gamma_k.
\]
Also, let $\setE = \cup_{k=1}^K \setE_k$, which is the event that the sufficient rank-one condition \eqref{eq:main_cond} in Theorem~\ref{thm:main} is violated.
Our problem is to prove an upper bound on ${\rm Pr}( \setE )$.
By the union bound, we have
\begin{equation} \label{eq:PrE_union}
{\rm Pr}( \setE ) \leq \sum_{k=1}^K {\rm Pr}( \setE_k ).
\end{equation}
Let us focus on ${\rm Pr}( \setE_k )$.
Under the Gaussian distribution assumption in Proposition~\ref{prop:prob}, it can be shown that
\[ 
\| \bbPi_k \bbh_k \|_2^2 = \frac{\rho_k}{2} U_{2(N-K+1)},
\]
where $U_d$ denotes a standard chi-square random variable with $d$ degrees of freedom;
see \cite[Chapter~8.3.1]{tse2005fundamentals}.
Moreover, it is known that
\[
{\rm Pr}( U_d \leq \beta d ) \leq \left( \beta e^{1-\beta} \right)^{\frac{d}{2}}, \quad
\text{for any $\beta \in (0,1)$;}
\]
see, e.g., \cite[Proposition A.4]{SY10}.
Using the above two results, we get
\begin{align*}
{\rm Pr}(\setE_k) & = {\rm Pr}\left(  U_{2(N-K+1)} \leq \frac{2\eps_k^2}{\rho_k} \left[ 1 + K + (K - \tfrac{1}{K} )  \gamma_k  \right] \right)  \\
& = {\rm Pr}\left( U_{2(N-K+1)} \leq 2(N-K+1) \frac{\eta_k}{{\sf CUR}_k} \right)    \\
& \leq \left( \frac{\eta_k}{{\sf CUR}_k} e^{1 - \frac{\eta_k}{{\sf CUR}_k} } \right)^{N-K+1},
\quad \text{for $\frac{\eta_k}{{\sf CUR}_k} < 1$}.
\end{align*}
By plugging the above inequality into \eqref{eq:PrE_union} and using $e^{1-\beta} \leq e$ for any $\beta \in (0,1)$, we obtain the desired result.


\subsection{Proof of Proposition~\ref{prop:wd2}}
\label{apx:wd}

The proof is by construction.
Recall the notation $\bbF= [~ \bbh_1, \ldots, \bbh_K ~]$.
Throughout the proof, we shall assume that
\begin{subequations}
\begin{align}
\bbF^H \bbF & = \bI, \quad N \geq K, \label{eq:wd2_setup_a}   \\
\sigma_1^2 & = \cdots = \sigma_K^2 \triangleq \sigma^2, 
\quad 
\gamma_1 
 = \cdots = \gamma_K \triangleq \gamma, \\
\eps_1 & = \cdots = \eps_K \triangleq \eps < 1.
\end{align}
\end{subequations}
The proof is divided into four steps.
In Step 1, we determine a condition on $\gamma, \eps$ under which Problem~\eqref{eq:main} has an optimal solution.
In Steps 2--3, we prove bounds on the LHS and RHS of \eqref{eq:wd}, respectively.
In Step 4, we combine the results in the previous steps and identify a case where \eqref{eq:wd}  has strict inequality 
for a particular setting of $N,K,\gamma,\eps$.

\medskip
{\em Step 1:} \
We show that Problem~\eqref{eq:main} has an optimal solution if 
\begin{equation} \label{eq:wd2_feas_cond}
\gamma < \frac{1}{K-1} \left( \frac{1}{\eps} - 1 \right)^2.
\end{equation}
Suppose that Problem~\eqref{eq:main} has a feasible solution $\tilde{\cW}$.
Since the optimal value $v^\star$ of Problem~\eqref{eq:main} satisfies 
$v^\star \leq \sum_i {\rm Tr}( \tilde{\bW}_i) < \infty$,
we can write
\begin{align*}
v^\star  = \inf_{ \cW \in \setS } & ~ \textstyle \sum_i {\rm Tr}(\bW_i)  \\
{\rm s.t.} & ~ \text{$\eqref{eq:main_b}-\eqref{eq:main_c}$,} ~ 
\textstyle \sum_i {\rm Tr}(\bW_i) \leq  R,
\end{align*}
where $R < \infty$ satisfies $\sum_i {\rm Tr}( \tilde{\bW}_i) < R$.
Since the above problem has a continuous objective function and a compact constraint set,
the optimal value $v^\star$ is attained. 

Our next question is when Problem~\eqref{eq:main} is feasible.
Let
\begin{equation*}
\bW_i = \alpha \bbh_i \bbh_i^H, ~ i=1,\ldots,K,
\end{equation*}
where $\alpha > 0$ is to be determined.
Putting the above $\cW$ into the LHS of \eqref{eq:main_b}, we see that
\ifconfver
    \begin{subequations}
    \begin{align}
    & \max_{ \bh_i \in \setU_i} \varphi_i(\cW,\bh_i)  =
    	\max_{ \| \be_i \|_2 \leq \eps } \varphi_i(\cW,\bbh_i+\be_i)  \\
    	& \quad  = \max_{ \| \be_i \|_2 \leq \eps } \sigma^2 + \alpha \left[  \sum_{j \neq i} | \be_i^H \bbh_j |^2 - \frac{1}{\gamma} | 1 + \be_i^H \bbh_i |^2    \right]  \label{eq:wd2_t1b} \\
    	& \quad \leq \sigma^2 + \alpha \left[  (K-1) \eps^2 - \frac{1}{\gamma} | 1 - \eps |^2    \right],  \label{eq:wd2_t1c}
    \end{align}
    \end{subequations}
\else
    \begin{subequations}
    \begin{align}
    \max_{ \bh_i \in \setU_i} \varphi_i(\cW,\bh_i) & =
    	\max_{ \| \be_i \|_2 \leq \eps } \varphi_i(\cW,\bbh_i+\be_i)  \\
    	& = \max_{ \| \be_i \|_2 \leq \eps } \sigma^2 + \alpha \left[  \sum_{j \neq i} | \be_i^H \bbh_j |^2 - \frac{1}{\gamma} | 1 + \be_i^H \bbh_i |^2    \right]  \label{eq:wd2_t1b} \\
    	& \leq \sigma^2 + \alpha \left[  (K-1) \eps^2 - \frac{1}{\gamma} | 1 - \eps |^2    \right],  \label{eq:wd2_t1c}
    \end{align}
    \end{subequations}
\fi
where \eqref{eq:wd2_t1b} is owing to \eqref{eq:wd2_setup_a}, and 
\eqref{eq:wd2_t1c} is obtained via standard vector inequalities.
Suppose that $(K-1) \eps^2 - \frac{1}{\gamma} | 1 - \eps |^2 < 0$, 
which can be easily verified to be equivalent to \eqref{eq:wd2_feas_cond}.
Then, from \eqref{eq:wd2_t1c}, we observe that the constraints in \eqref{eq:main_b} is satisfied if $\alpha$ is sufficiently large.
This implies that Problem~\eqref{eq:main} has a feasible solution if \eqref{eq:wd2_feas_cond} holds.

\medskip
{\em Step 2:} \
We show that the optimal value $v^\star$ of Problem~\eqref{eq:main}, or the problem on the LHS of \eqref{eq:wd}, has a lower bound
\begin{equation} \label{eq:wd2_v_lb}
v^\star \geq \frac{K \sigma^2}{ \frac{1}{\gamma} \left( 1 + \frac{\eps^2}{N} \right) - (K-1)  \frac{\eps^2}{N} }
\end{equation}
if $ \frac{1}{\gamma} (1 + \frac{\eps^2}{N} ) - (K-1)  \frac{\eps^2}{N} > 0$.
By the tight SDR representation of the robust constraints in 
\eqref{eq:setV} and \eqref{eq:main_sdr},
we have
\begin{align}
\max_{ \bh_i \in \setU_i} \varphi_i( \cW, \bh_i ) & = \max_{ \bH_i \in \setV_i} \phi_i( \cW, \bH_i ) \nonumber \\
& \geq \phi_i( \cW, \tilde{\bH}_i ) \quad \text{for any $\tilde{\bH}_i \in \setV_i$.}
\label{eq:wd2_t2}
\end{align}
Let us choose $\tilde{\bH}_i = \bbh_i \bbh_i^H + \frac{\eps^2}{N} \bI$, which can be verified to satisfy $\bH_i \in \setV_i$.
By letting $\alpha_i = {\rm Tr}(\bW_i) \geq 0$, we have
\begin{align}
\phi_i( \cW, \tilde{\bH}_i )&  \geq \sigma^2 - \frac{1}{\gamma} \bbh_i^H \bW_i \bbh_i 
	+ \frac{\eps^2}{N} \left( \sum_{j \neq i} \alpha_j - \frac{1}{\gamma} \alpha_i \right) \nonumber \\
	& \geq \sigma^2 -  \frac{1}{\gamma} \alpha_i
	+ \frac{\eps^2}{N} \left( \sum_{j \neq i} \alpha_j - \frac{1}{\gamma} \alpha_i \right),
	\label{eq:wd2_t3t}
\end{align}
where \eqref{eq:wd2_t3t} is due to $\bbh_i^H \bW_i \bbh_i  \leq \| \bbh_i \|_2^2 \cdot {\rm Tr}(\bW_i) = {\rm Tr}(\bW_i)$.
Applying \eqref{eq:wd2_t2}--\eqref{eq:wd2_t3t} to Problem~\eqref{eq:main} leads to the relaxation
\begin{subequations} \label{eq:wd2_t3}
\begin{align}
v^\star \geq \min_{ \bm\alpha \geq \bzero }  & ~ \textstyle \sum_i \alpha_i \label{eq:wd2_t3a} \\
{\rm s.t.} & ~ 
\textstyle 
\sigma^2 + \frac{\eps^2}{N} \sum_{j\neq i} \alpha_j - \frac{1 + \frac{\eps^2}{N}}{\gamma} \alpha_i \leq 0, ~i=1,\ldots,K.
\label{eq:wd2_t3b}
\end{align}
\end{subequations}
It is easy to show that for $ \frac{1}{\gamma} (1 + \frac{\eps^2}{N} ) - (K-1)  \frac{\eps^2}{N} > 0$,
the optimal solution $\bm\alpha^\star$ to Problem~\eqref{eq:wd2_t3} is\footnote{Concisely, we have $\alpha^\star_1 = \cdots = \alpha_K^\star$ because \eqref{eq:wd2_t3a} is Schur-convex and the constraints in \eqref{eq:wd2_t3b} do not depend on the ordering permutations of $\bm\alpha$. Consequently, we can reduce the problem to a single-variable problem, whose solution can be easily verified.}
\[
\alpha_i^\star = \frac{ \sigma^2 }{ \frac{1}{\gamma} \left( 1 + \frac{\eps^2}{N} \right) - (K-1)  \frac{\eps^2}{N} }, 
\quad i=1,\ldots,K.
\]
Substituting the above solution into \eqref{eq:wd2_t3}, we obtain the lower bound in \eqref{eq:wd2_v_lb}.

\medskip
{\em Step 3:} \
Let $d^\star$ denote the optimal value of the problem on the RHS of \eqref{eq:wd}.
We show that 
\begin{equation} \label{eq:wd2_d_ub}
d^\star \leq \frac{K \gamma \sigma^2}{ ( 1 - \sqrt{K} \eps )^2 }
\end{equation}
if 
$1 - \sqrt{K} \eps > 0$.
Let $\bF = [~ \bh_1, \ldots, \bh_K ~]$ and 
\begin{equation} \label{eq:wd2_t4}
\begin{aligned}
d(\bF) = \min_{ \cW \in \setS } & ~ \textstyle \sum_i {\rm Tr}(\bW_i) \\
{\rm s.t.}\, & ~ \varphi_i( \cW, \bh_i ) \leq 0, ~ \text{for all $i$}
\end{aligned}
\end{equation}
be the optimal value of the inner problem on the RHS of \eqref{eq:wd}.
Let 
\begin{equation} \label{eq:wd2_t5}
\bW_i = \alpha \bm u_i \bm u_i^H, \quad \bm u_i = \frac{ \bPi_i \bh_i }{ \| \bPi_i \bh_i  \|_2},
\quad i=1,\ldots,K,
\end{equation}
for some $\alpha > 0$,
where
$\bPi_i$ denotes the orthogonal complement projector of $\bF_{-i}$.
It will be seen that $ \| \bPi_i \bh_i  \|_2 > 0$ if 
$1 - \sqrt{K} \eps > 0$.
Substituting \eqref{eq:wd2_t5} into \eqref{eq:wd2_t4} yields
\begin{equation} \label{eq:wd2_t6}
\begin{aligned}
d(\bF) \leq \min_{ \alpha \geq 0 } & ~ \textstyle K \alpha \\
{\rm s.t.} & ~ \textstyle \sigma^2 - \frac{\alpha}{\gamma} \| \bPi_i \bh_i \|^2_2 \leq 0, ~ \text{for all $i$}.
\end{aligned}
\end{equation}
Using the result in Appendix~\ref{apx:eq:main_imp0},
we have
\begin{equation} \label{eq:wd2_t7}
\| \bPi_i \bh_i \|_2 \geq \sigma_{\rm min}(\bF).
\end{equation}
Also, by writing $\bF = \bbF + \bm E$ with $\bm E = [~ \be_1,\ldots, \be_K ~]$ and $\| \be_i \|_2 \leq \eps$ for all $i$,
and denoting $\sigma_{\rm max}(\cdot)$ as the largest singular value of its argument,
we obtain
\begin{align} 
\sigma_{\rm min}(\bF) & \geq \sigma_{\rm min}( \bbF ) - \sigma_{\rm max}( \bm E)  
\geq 1 - \sqrt{K} \eps,   \label{eq:wd2_t8}
\end{align}
where we have used the fact that $\sigma_{\rm max}( \bm E)^2 \leq {\rm Tr}(\bm E^H \bm E)$ and $ \sigma_{\rm min}( \bbF ) = 1$ (see \eqref{eq:wd2_setup_a}).
Eqs.~\eqref{eq:wd2_t7}--\eqref{eq:wd2_t8} imply that  $ \| \bPi_i \bh_i  \|_2 > 0$ if 
$1 - \sqrt{K} \eps > 0$.
Furthermore, applying \eqref{eq:wd2_t7}--\eqref{eq:wd2_t8} to \eqref{eq:wd2_t6} and assuming $1 - \sqrt{K} \eps > 0$ lead to
\begin{equation*} 
\begin{array}{r@{\hspace{4pt}}r@{\hspace{4pt}}l}
d(\bF) \leq & \displaystyle \min_{ \alpha \geq 0 } & ~ \textstyle K \alpha \\
&  {\rm s.t.} & ~ \textstyle \sigma^2  \leq \frac{\alpha}{\gamma} ( 1 - \sqrt{K} \eps )^2 \\
= &  \multicolumn{2}{l}{\displaystyle \frac{K \gamma \sigma^2}{ ( 1 - \sqrt{K} \eps )^2 }.}
\end{array}
\end{equation*}
Thus, we have \eqref{eq:wd2_d_ub}.

\medskip
{\em Step 4:} \
We combine the results in the previous steps to obtain the final result.
Steps 2--3 reveal that if 
\begin{equation} \label{eq:step4}
\frac{K \sigma^2}{ \frac{1}{\gamma} \left( 1 + \frac{\eps^2}{N} \right) - (K-1)  \frac{\eps^2}{N} } >
\frac{K \gamma \sigma^2}{ ( 1 - \sqrt{K} \eps )^2 }
\end{equation}
holds for some $(N,K,\gamma,\eps)$ such that 
$\frac{1}{\gamma} (1 + \frac{\eps^2}{N} ) - (K-1)  \frac{\eps^2}{N} > 0$ and
$1 - \sqrt{K} \eps > 0$, then $v^\star > d^\star$ holds.
Thus, our task is to construct such instances.
Assume $N \geq K \geq 5$,
\begin{equation} \label{eq:wd2_eps_setup} 
\eps = \frac{1}{2 N \sqrt{K} + 1},
\quad 
\gamma \leq \frac{4NK^2}{K-1}.
\end{equation}
It can be verified that $\frac{1}{\gamma} (1 + \frac{\eps^2}{N} ) - (K-1)  \frac{\eps^2}{N} > 0$ and
$1 - \sqrt{K} \eps > 0$.
Also, the condition in \eqref{eq:step4} can be rewritten as
\begin{equation} \label{eq:wd2_cond_g11}
\gamma > \frac{1}{K-1} ( 4N^2K - C ),
\end{equation}
where $C= NK - 2 N \sqrt{K} - 1$; note that $C > 0$ for $N \geq K \geq 5$.
Now, by choosing
\begin{equation}  \label{eq:wd2_gamma_setup} 
\gamma = \frac{1}{K-1} ( 4N^2 K - \delta ),
\end{equation}
for any $0 < \delta  < C$,
we see that \eqref{eq:wd2_cond_g11} is satisfied.
Hence, we have identified instances of $(N,K,\gamma,\eps)$  for which $v^\star > d^\star$.

We should also verify that the instances constructed above satisfy $v^\star < \infty$.
In Step 1, we showed that Problem~\eqref{eq:main} has an optimal solution, or $v^\star < \infty$, if \eqref{eq:wd2_feas_cond} holds.
For the choice of $\eps$ in  \eqref{eq:wd2_eps_setup},
it can be verified that \eqref{eq:wd2_feas_cond} becomes $$\gamma < \frac{1}{K-1} 4N^2 K. $$
As seen, the above condition is satisfied by the choice of $\gamma$ in \eqref{eq:wd2_gamma_setup}.
The proof is complete.

\bibliography{robust_unicast_ref,general_mimo}


\ifplainver
\else

\begin{IEEEbiography}[{\includegraphics[width=1in,height=1.25in,clip,keepaspectratio]{WingKinMa.eps}}]{Wing-Kin Ma}
(M'01-SM'11-F'17)
received the B.Eng. degree in electrical and electronic
engineering from the University of Portsmouth,
Portsmouth, U.K., in 1995, and the M.Phil. and
Ph.D. degrees, both in electronic engineering (EE), from
The Chinese University of Hong Kong (CUHK),
Hong Kong, in 1997 and 2001, respectively.
He is now an Associate Professor with the
Department of EE, CUHK. 
His research interests are in signal processing, 
communications and optimization.
He
currently serves as Senior Area Editor of {\sc IEEE Transactions on Signal Processing} and Associate Editor of {\em Signal Processing}, and he is  a member of the Signal Processing Theory and Methods
 Technical Committee (TC)
and the Signal Processing for Communications and Networking TC.
He received
2013--2014 CUHK Research Excellence Award,
the 2015 IEEE Signal Processing Magazine Best Paper Award,
and the 2016 IEEE Signal Processing Letters Best Paper Award.
\end{IEEEbiography}

\vspace{-1em}

\begin{IEEEbiography}[{\resizebox{.92in}{!}{\includegraphics{JiaxianPan.eps}}}]
{\bf Jiaxian Pan} received the B.Eng. degree  from Sun Yat-sen (Zhongshan) University, Guangzhou, China, in 2008
and the Ph.D. degree in electronic engineering from the Chinese University of Hong Kong (CUHK), Hong Kong, in 2014.
He was a research associate in CUHK from 2014 to 2015.
He is currently working as a Senior Engineer with MediaTek Inc., Hsinchu, Taiwan.
Dr. Jiaxian Pan was awarded a Best Student Paper Award at ICASSP 2011.
\end{IEEEbiography}

\begin{IEEEbiography}[{\includegraphics[width=1in,height=1.25in,clip,keepaspectratio]{AnthonySo.eps}}]{Anthony Man-Cho So}
(M'12) received his PhD degree in Computer Science with a PhD minor in Mathematics from Stanford University.  Dr. So joined The Chinese University of Hong Kong (CUHK) in 2007.  He currently serves as Assistant Dean of the Faculty of Engineering and is an Associate Professor in the Department of Systems Engineering and Engineering Management.  He also holds a courtesy appointment as Associate Professor in the CUHK-BGI Innovation Institute of Trans-omics.

Dr. So currently serves on the editorial boards of {\scshape IEEE Transactions on Signal Processing}, Journal of Global Optimization, Optimization Methods and Software, and SIAM Journal on Optimization. He received the 2015 IEEE Signal Processing Society Signal Processing Magazine Best Paper Award, the 2014 IEEE Communications Society Asia-Pacific Outstanding Paper Award, the 2010 Institute for Operations Research and the Management Sciences (INFORMS) Optimization Society Optimization Prize for Young Researchers, and the 2010 CUHK Young Researcher Award. 

\end{IEEEbiography}

\begin{IEEEbiography}[{\includegraphics[width=1in,height=1.25in,clip,keepaspectratio]{TsungHuiChang.eps}}]{Tsung-Hui Chang}
(S'07--M'08) 
received the B.S. degree in electrical engineering and the Ph.D. degree in communications engineering from the National Tsing Hua University (NTHU), Hsinchu, Taiwan, in 2003 and 2008, respectively. He is currently an Assistant Professor with the School of Science and Engineering, The Chinese University of Hong Kong, Shenzhen, China. His research interests include signal processing and optimization problems in data communications and machine learning. Dr. Chang received the Young Scholar Research Award of National Taiwan University of Science and Technology (NTUST) in 2014, IEEE ComSoc Asian-Pacific Outstanding Young Researcher Award in 2015, and The Thousand Youth Talent Program of China in 2016. He currently serves as an Associate Editor of {\sc IEEE Transactions on Signal Processing} and {\sc IEEE Transactions on Signal and Information Processing over Networks.}

%
\end{IEEEbiography}

\fi

\end{document}